\documentclass{article}
\usepackage[bottom=1in,top=1in,left=0.5in,right=0.5in]{geometry}
\usepackage{graphicx}
\usepackage{cite}
\usepackage{newtxtext}
\usepackage{newtxmath}
\usepackage{natbib}
\usepackage{hyperref}

\usepackage{tikz}

\usepackage{multirow}
\usepackage[colorinlistoftodos]{todonotes}

\hypersetup{
    colorlinks = true,
    urlcolor   = blue,
    citecolor  = blue,
    linkcolor=blue,
}

\newcommand{\RomanNumeralCaps}[1]
\linenumbers

\definecolor{my_green}{HTML}{009051}

\title{Suppressing instabilities in mixed baroclinic flow using an actuation based on receptivity}

\author{Abhishek Kumar$^{1,}$\thanks{Email: abhishek.kir@gmail.com}\,\, and  Alban Poth{\'e}rat$^{1}$\\
 Centre for Fluid and Complex Systems, Coventry University, Coventry CV1 5FB, United Kingdom}
 
\date{}

\begin{document}
\maketitle
%
\abstract{This paper presents a method to stabilise oscillations occurring in a mixed convective flow in a nearly hemispherical cavity, using actuation based on the receptivity map of the unstable mode. This configuration models the continuous casting of metallic alloys, where hot liquid metal is poured at the top of a hot sump with cold walls pulled in a solid phase at the bottom. The model focuses on the underlying fundamental thermo-hydrodynamic processes without dealing with the complexity inherent to the real configuration \citep{Flood:MST1994}. This flow exhibits three branches of instability \citep{Kumar:JFM2020}.  The solution of the adjoint eigenvalue problem for the convective flow equations reveals that the regions of highest receptivity for unstable modes of each branch concentrate near the inflow upper surface. Simulations of the linearised governing equations {show that a thermo-mechanical actuation modelled on the adjoint eigenmode asymptotically suppresses the unstable mode. If the actuation's amplitude is kept constant in time, which is easier to implement in an industrial environment, the suppression is still effective but only} over a finite time, after which it becomes destabilising. Based on this phenomenology, we apply the same actuation during the stabilising phase only in the nonlinear evolution of the unstable mode. It turns out stabilisation persists, even when the unstable mode is left to evolve freely after the actuation period. These results not only demonstrate the effectiveness of receptivity-informed actuation in stabilising convective oscillations but also suggest a simple strategy for their long-term control.}
%
%
\section{Introduction}
\label{sec:intro}
This paper is concerned with the suppression of oscillations in mixed convective flows in an open cavity permeated by a through-flow. It is inspired by the occurrence of such oscillations during the continuous casting of liquid metal alloys. In this type of process, solidified metal is pulled from the bottom of a pool of melted metal continuously fed from above. The pulling speed is adjusted to match that of the solidification front which, therefore, behaves as a steady but porous boundary for the fluid.  A key issue is the appearance of oscillations resulting in unwanted macro-segregations~\citep{Dorward:Al,Beckermann:Macrosegregation} near the axis of the melt.  We previously showed that the mechanism underpinning these oscillations could be reproduced in a simple hemispherical model of the sump \citep{Flood:MST1994} capturing the rather unusual interplay between the baroclinic convection caused by the cooled lateral wall \citep{pierrehumbert1995_arfm} and the through-flow \citep{Kumar:JFM2020}. Despite ignoring the complexities of the chemistry, multiple phase and solidification of the real process \citep{kuznetsov1997_cht,sheng2000_mmtb,thomas2001_isiji}, the model produced three branches of linear instability: one supercritical and oscillatory, one subcritical and oscillatory, and one supercritical and non-oscillatory when the Reynolds number $Re=u_0 H/\nu$ based on the inflow velocity $u_0$, sump height $H$ and fluid kinematic viscosity $\nu$ was varied. The topology of the  oscillatory modes, consistent with observations in the casting process points to their hydrodynamic nature, and so validates the simplified hydrodynamic approach. The purpose of this paper is to take further advantage of the mathematical tractability of this model to design an actuation capable of suppressing these instabilities.

The idea of suppressing oscillations by means of an actuator producing oscillations at a well-chosen location has been long-exploited in thermoacoustics \citep{lieuwen2003_ti, noiray2009_pci, dowling2005_arfm}, in particular to control diffusion flames in combustion \citep{magri2013_jfm,magri2014_jfm,juniper2018_arfm}. However, it is yet to make its way to metallurgy, where oscillations occur on much larger timescales. Yet, both problems share similar challenges: The design of an effective control loop requires not only a sufficiently accurate model of the system's dynamics but also sensors and actuators capable of feeding in the controller and enacting its output onto the process. The high temperatures, the highly corrosive nature of liquid metals and the risk of alloy contamination precludes the long term immersion of any such device in the melt. Hence, just like in combustion problems, their implementation in hostile environments often proves impractical or unreliable \citep{mongia2003_jpp}, so in both cases, open-loop control using a single actuator is preferred for its simplicity and robustness. While these techniques have long been implemented in thermoacoustics with actuators placed within the flow~\citep{mcmanus1990_cf,lubarsky2003_aiaa}, using them in liquid metals requires their positioning at the flow boundary. The key challenge is to find an actuation satisfying these conditions.

A possible solution lies in the idea of structural sensitivity, best voiced in \cite{luchini2014_arfm}'s review on adjoint equations in stability problems:
\emph{``The key reasoning is that, if indeed a specific spatially localised region (a wave maker) acts as the driver of the oscillation and the rest of the flow just amplifies it, a structural perturbation acting in the amplifier portion is bound to affect only the amplitude (eigenvector) and not the frequency (eigenvalue) of the oscillation. Conversely, a perturbation in the wave-maker region mostly affects the eigenvalue. The structural sensitivity of the eigenvalue thus acts as a marker for the spatial location of the wave maker.''}  For the problem we are considering structural sensitivity thus offers a method to calculate the position and topology of the actuation best suited to suppress the growth of the unstable mode underpinning the onset of oscillations. It is all the better suited as we already identified the unstable modes in a previous linear stability analysis performed for each of the three branches appearing at different Reynolds numbers~\citep{Kumar:JFM2020}.

Indeed, structural sensitivity has successfully informed design alterations in devices where oscillations driven by instabilities took place, as in combustion problems or in the recent example of the stabilisation of inkjet in printers \citep{aguilar2020_prf,kungurtsev2018_jfm}. The other common application of this approach concerns the \emph{a posteriori} identification of suppression mechanisms where existing actuators are already implemented: the long misunderstood suppression of the von K\'arm\'an street by a small control cylinder placed in the near wake \citep{strykowski1990_jfm}, was thus successfully explained when \cite{marquet2008_jfm} were able to account for the feedback of the cylinder on both the unstable perturbation and the base flow. Further, a similar approach has been used in studying boundary-layer stability~\citep{Park:JFM2019,Brandt:JFM2011}. Here we consider a  different approach aligned with the industrial requirements of finding actuators best suited to suppress the oscillations and adaptable when the flow parameters are varied (here the Reynolds number, for example). 
Since such an actuator acts either mechanically or thermally on the system, we model this combined actuation by combining an external force in the momentum equations and an external source term in the energy equation. {Since the adjoint mode corresponds to the Green function for the receptivity to an external actuation, aligning the forcing on it offers a way to directly control the amplitude of the unstable mode (see equation (3.10) from \cite{GIANNETTI:JFM2007}). This approach follows a different principle from those based on base-flow sensitivity, as developed by \citet{marquet2008_jfm}. In these, the actuation aims to alter the base flow to shift the eigenvalue of the unstable mode towards the stable region. Hence, the idea is to shift the system so that it becomes stable. In our approach, by contrast, we do not alter the base flow or any other part of the base system but add an actuation that targets only the unstable mode once it grows. In the strategy using base-flow sensitivity, the actuation is optimised to alter the base flow, whereas in our receptivity-based strategy, the actuation must leave the base flow unaffected and act only on the perturbation. Doing so, however, raises three difficulties. First, we must ensure that the base flow is sufficiently unaffected by the actuation for the unstable mode to retain the topology targeted by the actuation. This can be verified using fully nonlinear simulations of the actuated system. Second,} the linear theory does not provide an indication of the amplitude or the phase of the forcing, neither relative to that of the unstable mode, nor absolute. Both parameters therefore need to be varied to find the most effective combination for the suppression of the oscillations. Third, the system may evolve out of the linear regime where structural sensitivity operates. At this point, the system's evolution ceases to follow that for which the forcing was designed in the first place. In control language, the system does not follow the controller's model, so further applying the forcing may not result in the suppression predicted by the linear model. Hence the actuation may only be effective for a finite time. The question is whether this time is sufficiently long for any meaningful suppression strategy based on this approach.

We propose to answer these questions by performing the receptivity and sensitivity analyses based on the stability analysis we conducted on the mixed convective flow in a cavity in \cite{Kumar:JFM2020} and numerically apply an external forcing built as described above. Besides exploring the idea of instability suppression by receptivity-informed external force, this problem carries three specificities of further fundamental interest from the physical and mathematical point of view: First, the mixed-convection character of this flow combines an open flow, for which structural sensitivity analysis has been perhaps most developed~\citep{GIANNETTI:JFM2007,GIANNETTI:JFM2010}, with a buoyant flow, for which, to our knowledge, it was only used on stratified wakes~\citep{Chen:JFM2017}. Aside of this example, open-loop control \citep{tang1998_pf,bau1999_ijhmt} and stabilisation by vibrating walls \citep{anilkumar1993_jap,medelfef2023_pf} have been implemented in classical Rayleigh-B\'enard and Marangoni flows. The adjoint equations for convective flows also made it possible to determine the influence of specific temperature distributions and heat fluxes at the flow boundaries (see \cite{momose1999_jsmeb} and others) and to infer past states of the Earth's mantle \citep{bunge2003_gji, ismail-zadeh2004_pepi, horbach2014_ijg}. Combining adjoint equations with linear stability analysis for the purpose of identifying the sources of convective instabilities and suppressing them, however, presents a new and interesting mathematical problem.

Convective flows are indeed particularly interesting in this context, as in most cases, they occur in combination with other effects such as shear flows in mixed convection (as in \cite{vo2017_prf} and in the present case), or Lorentz and Coriolis forces in the vast field studying liquid planetary interiors \citep{roberts2013_prp}. Because of this, the path to instability may follow different branches of instability, either individually near the onset or simultaneously in more supercritical regimes, leading to multi-modality \citep{nakagawa1957_prsa,eltayeb1972_prsa,aujogue2015_pf,horn2022_prsa,horn2023_prsa,xu2023_prf}. For this reason, suppressing instabilities may require different actuations for different branches. These may even be used in combination in multi-modal regimes, although their effectiveness may be limited if nonlinear interactions between these modes become significant. Whether the problem of mixed convection in a cavity may become multi-modal when sufficiently supercritical is, at this point, an open question. For this reason, {having} in mind the aim of exploring the feasibility of suppressing convective oscillations using receptivity-informed actuation, we shall restrict ourselves to weakly supercritical regimes where instabilities are driven by a single unstable mode in each of the three branches of instability.  This still leaves  the question open as to whether the approach would be equally effective for each of them, especially so as these occur through bifurcations of different nature. Hence, we shall seek answers to the following questions:
\begin{itemize}
\item[(i)] Does the system have significant receptivity in regions of the flow where an actuation can realistically be applied, in particular near the boundaries?
\item[(ii)] {Which forcing parameters (phase and amplitude) are best suited for applying an external thermo-mechanical actuation, modelled on the adjoint of the eigenmode, to achieve suppression?}
\item[(iii)] By how much can the energy of the oscillations be reduced and for how long?
\item[(iv)] Does an actuation purely based on linear dynamics remain effective when the nonlinearities are accounted for?
\end{itemize}
To answer these questions, we start by formulating the adjoint problem for mixed baroclinic convection in a nearly-hemispherical cavity, as defined by \cite{Kumar:JFM2020} and recalled in \S~\ref{sec:formulation}, along with a description of the numerical system based on high-order spectral elements methods. We then identify the thermo-mechanical source of the instabilities by performing receptivity and sensitivity analyses (\S~\ref{sec:receptivity_sensitivity_anal}). To determine the optimal phase and amplitude of the actuation based on the adjoint mode, relative to the unstable mode, we evolve the linearised equations, varying these values (\S~\ref{sec:Linear_receptivity_forcing}). Finally, we put the idea to the test in fully nonlinear simulations and assess how long nonlinearities are kept at bay (\S~\ref{sec:Nonlinear_receptivity_forcing}). 

\section{Problem formulation}
\label{sec:formulation}
\subsection{Governing equations}
\begin{figure}
\begin{center}
\includegraphics[width=0.5\textwidth]{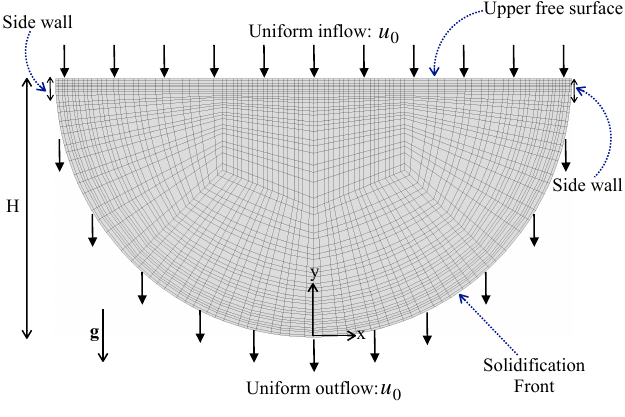}
\end{center}
	\caption{Problem geometry and mesh, with a rigid free surface at the inlet (top), solid side walls, and a porous solid wall at the outlet (bottom, solidification front). The flow enters and leaves the domains at vertical velocity $u_0$. The 
sketch also shows details of the mesh. This mesh consists of  $348$ quadrilateral elements, and each element is represented by the polynomial order of $N=3$. Thus, the total collocation points are $348 \times (N+1)^2 = 5568$.} 
\label{fig:mesh_flow}
\end{figure}

Following \cite{Kumar:JFM2020}, we consider a cavity of height $H$ with an upper free surface where hot fluid is fed in, and a cold, porous lower boundary representing a solidification front, as shown in figure~\ref{fig:mesh_flow}. The cavity is also assumed infinitely extended in the third direction ($\mathbf e_z$). The lower boundary is semicircular and connected to the flat upper boundary by two solid, adiabatic side walls of height $0.05H$, with a temperature difference $\Delta T$ between top and bottom boundaries. To model a liquid metal of density $\rho$, kinematic viscosity $\nu$, thermal diffusivity $\alpha$, thermal expansion coefficient $\beta$, the flow is assumed Newtonian and since temperature gradients remain moderate, its motion is assumed well described by the Oberbeck--Boussinesq approximation~\citep{Oberbeck:1879, Boussinesq:book, Chandrasekhar:book}. This leads to the following non-dimensional governing equations:
\begin{eqnarray}
\frac{\partial {\bf u}}{\partial t}+ ({\bf u} \cdot \nabla) {\bf u} + \nabla p & = & RaPr T {\bf e}_y+Pr \nabla^2 {\bf u},\label{eq:u}  \\
\frac{\partial T}{\partial t}+ ({\bf u} \cdot \nabla) {T} & = &  \nabla^2 T, \label{eq:T} \\
\nabla \cdot {\bf u} & = & 0\label{eq:div_u} , 
\end{eqnarray}
where ${\bf u}=(u,v,w)$ is the velocity vector field,  $T$ is the temperature field, $t$  is the time, and ${\mathbf g=-g\mathbf e_y}$ is the gravitational  acceleration. The modified pressure $p$ includes the buoyancy term that accounts for the reference temperature $T_0$ at density $\rho_0$~\citep{Chandrasekhar:book,Tritton:book}. The above set of equations are non-dimensionalised using length $H$, velocity $\alpha/H$, time $H^2/\alpha$, pressure $\rho_0(\alpha/H)^2$ and temperature $\Delta T$. The equations feature two governing non-dimensional groups: The Prandtl number 
\begin{equation}
Pr  = \frac{\nu}{\alpha},
\end{equation}
which we fixed to $0.02$, a typical value of aluminium alloys, and the Rayleigh number $Ra$, defined as
\begin{equation}
Ra = \frac{\beta g \Delta T H^3}{\nu \alpha}, \label{eq:Ra}
\end{equation}
which controls the intensity of buoyancy forces relative to viscous forces. 
A free-slip boundary condition is applied to the upper boundary at $y=1$ with fluid being poured homogeneously across the boundary at an imposed temperature $\Delta T$. These boundary conditions are expressed as:
\begin{eqnarray}
\frac{\partial }{\partial y} \mathbf u\times \mathbf e_y=0, \qquad \mathbf u\cdot \mathbf e_y=RePr, \qquad 
T(y=1)=1,
\label{eq:bc_top}
\end{eqnarray}
where $Re$ is the mass flux Reynolds number based on the dimensional feeding velocity $u_0$  
\begin{equation}
Re = \frac{u_0H}{\nu}.
\end{equation}
At the lower boundary $\mathcal S$, the solidification front is represented by solid, porous boundary conditions:
\begin{eqnarray}
\mathbf u_{\mathcal{S}}\times \mathbf e_y=0, \qquad \mathbf u_{\mathcal{S}}\cdot \mathbf e_y=RePr, \qquad T_{\mathcal{S}}=0,
\label{eq:bc_bot}
\end{eqnarray}
such that the flux of fluid pulled at $\mathcal{S}$ exactly cancels the flux of mass coming from the inlet. Impermeable, no-slip boundary conditions for the velocity field and insulating boundary conditions for the temperature field are imposed at the side-walls (see figure~\ref{fig:mesh_flow}). These boundary conditions for the temperature field at these side-walls ensure that the temperature field remains continuous along the entire periphery of the domain. {The pressure field in our calculation is obtained by solving the Poisson equation derived by taking the divergence of Eq. (\ref{eq:u}). A consistent Neumann boundary condition for pressure~\citep{Karniadakis:JCP1991} is applied at $y = 1$, on the lower boundary $\mathcal{S}$, and on the side-walls.} In the third direction ${\bf e}_z$, the domain's infinite extension is represented by periodic boundary conditions for all flow fields.

\subsection{Direct and adjoint perturbation equations}
\label{ssec:direct_adjoint_eqn}
The purpose of this work is to suppress linear instabilities with an actuation specifically designed to stifle the growth of linear perturbations. Since this actuation will be based on the linear dynamics of this perturbation, 
we first need the establish the set of equations that govern its linear dynamics. From \cite{Kumar:JFM2020}, linear perturbations grow 
from a class of equilibrium solutions of Eqs. (\ref{eq:u}-\ref{eq:div_u}) and boundary conditions (\ref{eq:bc_top},~\ref{eq:bc_bot}) that are planar and invariant along ${\bf e}_z$, hence of the form ${\bf U}(x,y)$, $\bar{T}(x,y)$. Baroclinicity due to the isothermal condition along the solidification front precludes any purely diffusive thermal equilibrium, so ${\bf U}(x,y)$ is never homogeneously zero and both ${\bf U}(x,y)$ and $\bar{T}(x,y)$  must be found as a fully nonlinear two-dimensional solution of the equations. 
It follows that perturbations to this equilibrium ${\bf q}^\prime (x,y,z,t)=({\bf u},T)^\top - ({\bf U}, \bar{T})^\top=({\bf u}^\prime(x,y,z,t), T^\prime(x,y,z,t))^\top$, are governed by the linear system of equations governing the evolution of  infinitesimal perturbations,
\begin{equation}
\frac{\partial {\bf q}^\prime}{\partial t}  = \mathcal{L}  {\bf q}^\prime \label{eq:direct},
\end{equation}
where
\begin{equation}
\mathcal{L}{\bf q}^\prime
=
\begin{pmatrix}
    - ({\bf U} \cdot\nabla)  {\bf u^\prime}-( \nabla {\bf U}) \cdot {\bf u^\prime} -\nabla p^\prime + RaPrT^\prime {\bf e}_y+Pr{\nabla}^2{\bf u}^\prime    \\
    -{\bf U} \cdot\nabla T^\prime  -  (\nabla\bar{T})\cdot {\bf u}^\prime   + \nabla^2 T^\prime \\
\end{pmatrix}. \label{eq:L_opt}
\end{equation}
{Since $p^\prime$ is determined by the constraint $\nabla \cdot {\bf u}^\prime = 0$, it is therefore not included in the state vector ${\bf q}^\prime$.} We will refer to equation (\ref{eq:direct}) as the direct perturbation equation, and to $\mathcal{L}$ as the direct linear operator. The boundary conditions for the base flow are the same as those for the main variables. As a result, the perturbation variables satisfy the homogeneous counterparts of the boundary conditions associated with the base flow. Since the base flow is invariant along ${\bf e}_z$ and $\mathcal{L}$ does not explicitly depend upon time, the perturbation can be written as a linear combination of normal modes of the form: 
\begin{equation}
{\bf q}^\prime(x,y,z,t) = \hat{\bf q}(x,y) e^{ikz + \lambda t},
	\label{eq:normal_mode}
\end{equation}
where $k$ is the wavenumber along the homogenous direction ${\bf e}_z$ and $\lambda = \sigma \pm i\omega$ contains the growth rate, $\sigma$ and frequency $\omega$. The growth rate, frequency, and the wavenumber of the most dominant mode $\hat{\bf q}(x,y)=(\hat{u}, \hat{v}, \hat{w}, \hat{T})^\top$ (also referred to as the direct mode) are found by solving the eigenvalue problem for $\lambda$ that result from (\ref{eq:direct}) and (\ref{eq:normal_mode}). This is done numerically by means of a time-stepper method~\citep{Barkley:IJNMF2008}. Here the eigenmodes are normalised such that $\|\hat{\bf q}\|_2=1$, where $\|\cdot\|_2$ denotes the standard $l^2$ vector norm of  $4 \times N_e \times (N+1)^2$ values that make up $\hat{\bf q}$. In this context, $N_e$ refers to the number of quadrilateral elements. For the details of the eigenvalue solver we refer the reader to section~2.2 of our previous work \citep{Kumar:JFM2020}.

Next, we need to work out the form of the actuation that is best suited to suppress the growth of individual normal modes. The idea we pursue relies on  \cite{GIANNETTI:JFM2007}'s ideas, who showed that the receptive 
regions of the direct linear modes are mapped by the adjoint eigenmodes of the same linearised equations. We, therefore, need to construct the adjoint operator 
$\mathcal{L}^*$ with respect to the time-averaged inner product relevant to the problem~\citep{Mao:JFM2015}:
\begin{equation}
({\bf a}, {\bf b}) = \int_\Omega {\bf a}\cdot {\bf b} \, dV, \qquad \langle {\bf c},{\bf d} \rangle = \int_0^\tau ({\bf c} , {\bf d}) \, dt,
\end{equation}
where ${\bf a}$ and ${\bf b}$ are time-averaged vector  fields  defined  on the fluid domain $\Omega$, while ${\bf c} $ and ${\bf d}$ are  time-dependent vector  fields  defined  on $\Omega$ and  time  domain $[0, \tau]$. The adjoint operator is then defined by the relation 
\begin{equation}
   \langle{\bf q}^*  , (-\partial_t +\mathcal{L}){\bf q}^\prime \rangle- \langle (\partial_t+ \mathcal{L}^*){\bf q}^*, {\bf q}^\prime \rangle=0, \label{eq:adjoint_condition}
\end{equation} 
and thus satisfies:
\begin{equation}
-\frac{\partial {\bf q}^*}{\partial t} =   \mathcal{L}^*   {\bf q}^* \label{eq:adjoint},
\end{equation}
where ${\bf q}^*(x,y,z,t) = (u^*, v^*, w^*, T^*)^\top$ represents adjoint variables. The expression of $\mathcal L^*$ is readily derived by applying integration by parts to the first term in equation (\ref{eq:adjoint_condition}):
\begin{equation}
\mathcal{L}^*{\bf q}^*
=
\begin{pmatrix}
    ({\bf U} \cdot\nabla) {\bf u^*} - (\nabla {\bf U})^\top\cdot {\bf u^*}  -(\nabla \bar{T})T^*  -\nabla p^* +  Pr{\nabla}^2{\bf u}^*    \\
  ({\bf U} \cdot \nabla) {T^*}+ RaPr\, v^*+\nabla^2{T}^*, \\
\end{pmatrix}
\label{eq:L_start_opt}
\end{equation}
with $\nabla \cdot {\bf u}^*  = 0$. 
Similarly, the adjoint variables satisfy adjoint boundary conditions imposed by equation~(\ref{eq:adjoint_condition}). These are identical to the boundary conditions 
satisfied by the direct variables, except for the ${\bf e}_x$ and ${\bf e}_z$ components of the velocity field at the upper free surface, that must satisfy Robin conditions~\citep{Barkley:IJNMF2008}: 
\begin{equation}
{\bf e}_y \cdot \nabla u^* - Re \, u^*=0; \qquad {\bf e}_y \cdot \nabla w^* -Re \, w^*=0.
\end{equation}
Like the direct variables, the adjoint variables are decomposed into the normal modes but obtained as the solution of the adjoint eigenvalue problem, instead of the direct one so this time, the same eigenvalue solver yields the adjoint mode $\hat{\bf q}^*(x,y)=(\hat{u}^*, \hat{v}^*, \hat{w}^*, \hat{T}^*)^\top$. {It follows from the bi-orthogonality property that the eigenvalue of the adjoint mode is the complex conjugate of that of the direct mode~\citep{Salwen:JFM1981}. Therefore, the magnitude of the growth rate and frequency of the adjoint and direct modes are identical when the real and imaginary parts of the complex eigenvalue are equated.}

\subsection{Receptivity and sensitivity}
\label{ssec:RSA_formulation}
\cite{GIANNETTI:JFM2007} showed that the adjoint field represented Green’s function for the receptivity problem. Therefore, the adjoint equations can be used to evaluate the effects of any external {actuation or a generic initial condition on the leading eigenmode of the direct problem (\ref{eq:direct}).} This property forms the basis for the analysis in \S~\ref{sec:Linear_receptivity_forcing}, where we seek to identify the regions where applying an actuation would most effectively affect the growth of unstable modes. Our intention is to control the growth of the perturbation from an initial condition where the unstable mode has already grown measurably, but sufficiently little to remain within the confines of the linear approximation. For this purpose, we determine the topology of the actuation on the adjoint eigenmode. A common alternative approach is to force the system in such a way as to modify the underlying operator such that the eigenvalue associated with the leading eigenmode remains in the stable region, ignoring the initial condition. The optimal actuation for this purpose is given by the sensitivity map. The relative shift in the eigenvalue, and therefore the growth rate associated to the direct and the adjoint mode incurred by acting at any given location of the flow is given by the sensitivity map~\citep{GIANNETTI:JFM2007,GIANNETTI:JFM2010,Qadri:JFM2013}
\begin{equation}
S_{ij}(x,y)=\frac{\hat{\bf q}_i \hat{\bf q}^*_j }{\int_\Omega \hat{\bf q}^\top \hat{\bf q}^* \, dxdy}.
\end{equation}
Note that the sensitivity tensor above is based on feedback localised in space, of the form ${\mathbf C_0} \delta({\mathbf x} - {\mathbf x_0}) \hat{\bf q}$. Here, ${\mathbf C_0}$ denotes a constant coefficient matrix, $\mathbf x_0$ indicates the position where the feedback acts, and $\delta({\mathbf x} - {\mathbf x_0})$ denotes the Dirac delta function.Tensor $S_{ij}$ determines the relative local intensity of the feedback of individual component $\hat{\bf q}^*_j$  onto the individual component $\hat{\bf q}_i$ of the eigenmode. This quantity also locates the regions of the flow acting as {\em wavemakers}. In particular, since ${\mathbf q}^\prime$ and $\mathbf q^*$ contain 
all three components of velocity and the temperature, the knowledge of $S_{ij}$ indicates whether the actuation should be 
of thermal or mechanical nature and if mechanical, which component of the velocity (or combination of all four components including temperature) is most efficient at altering the growth of the unstable mode.
Both the direct and adjoint problems being linear, amplitudes are relative so we may further normalise the direct and adjoint modes by choosing~\citep{GIANNETTI:JFM2007}:
\begin{equation}
\int_\Omega \hat{\bf q}^\top  \hat{\bf q}^* \, dxdy = 1,\label{eq:normalisation_condition}
\end{equation}
so that the sensitivity tensor is simply expressed as $S_{ij}=\hat{\bf q}_i \hat{\bf q}^*_j$. 

At this point, we reiterate that our study considers the feedback on the perturbed field. This differs from \cite{marquet2008_jfm}'s approach, which examines the effect of forcing on the base flow. The objective of our study is to apply a forcing based on the adjoint mode (receptivity) to suppress instability. In general, the forcing influences both the base flow and the perturbation and this effect can be analysed in detail by studying the sensitivity to base flow modification as ~\citet{marquet2008_jfm} do or by validating the approach using nonlinear DNS as we do in this paper.  To summarise the difference between these two strategies in a nutshell, our strategy consists in applying a forcing that suppresses the instability \emph{before} either the perturbation or the forcing has sufficiently grown to affect the base flow. The strategy proposed by \citet{marquet2008_jfm}, by contrast, is to modify the base flow to prevent the perturbation from growing at all. 

\subsection{Methodology and choice of parameters}
\label{ssec:Methodology_parameters}
\begin{figure}
\begin{center}
\begin{tikzpicture}[scale = 0.8, every node/.style={transform shape}]
\node [below] at (-2.8,0.45) {(a)};
\node [below, xshift=100] at (3.2,0.45) {(b)};
\node [below, yshift=-30] at (-2.8,-3.3) {(c)};
\node [below, xshift=100, yshift=-30] at (3.2,-3.3) {(d)};
\node [below] at (0.2,0.45) {$Re=0;\,Ra=7\times 10^3$};
\node [below,  xshift=100] at (6.2,0.45) {$Re=50;\,Ra=7\times 10^3$};
\node [below,  yshift=-30] at (0.2,-3.3) {$Re=100;\,Ra=4\times 10^4$};
\node [below,  xshift=100, yshift=-30] at (6.2,-3.3) {$Re=150;\,Ra=8\times 10^4$};

\node[anchor=north, inner sep=0] (image) at (0,0) {\includegraphics[width=0.44\textwidth]{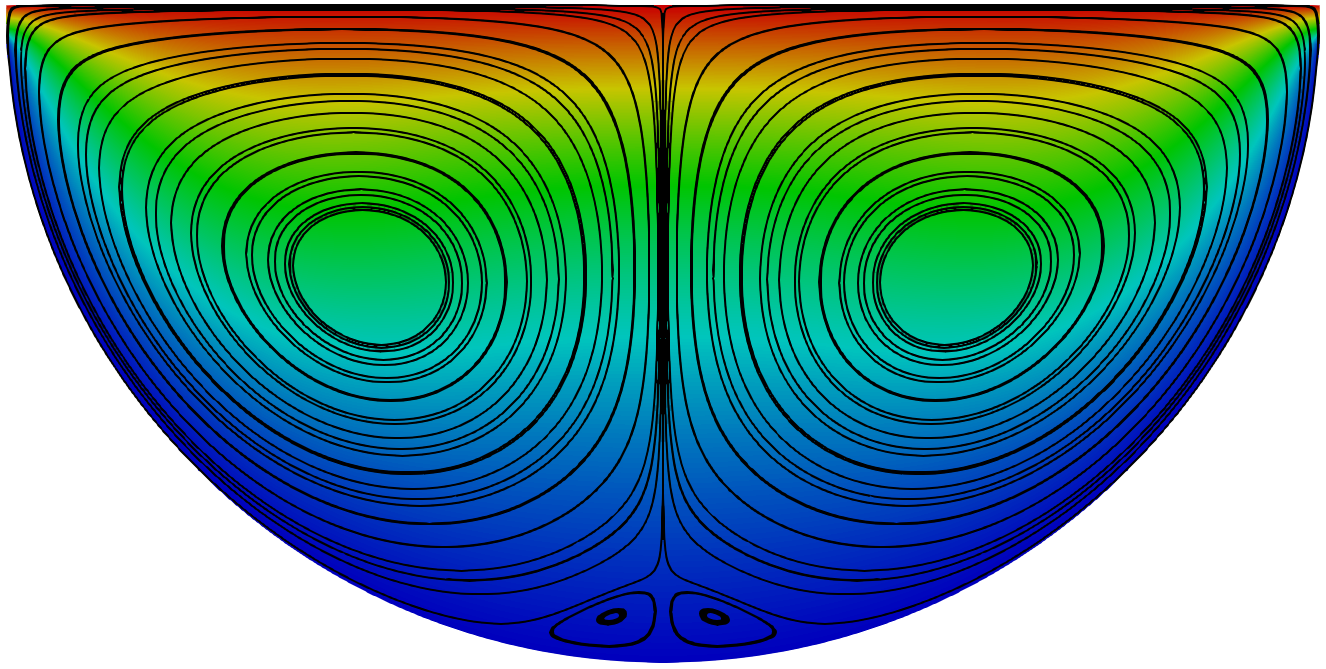}};
\node[anchor=north, inner sep=0, xshift=100] (image) at (6,0) {\includegraphics[width=0.44\textwidth]{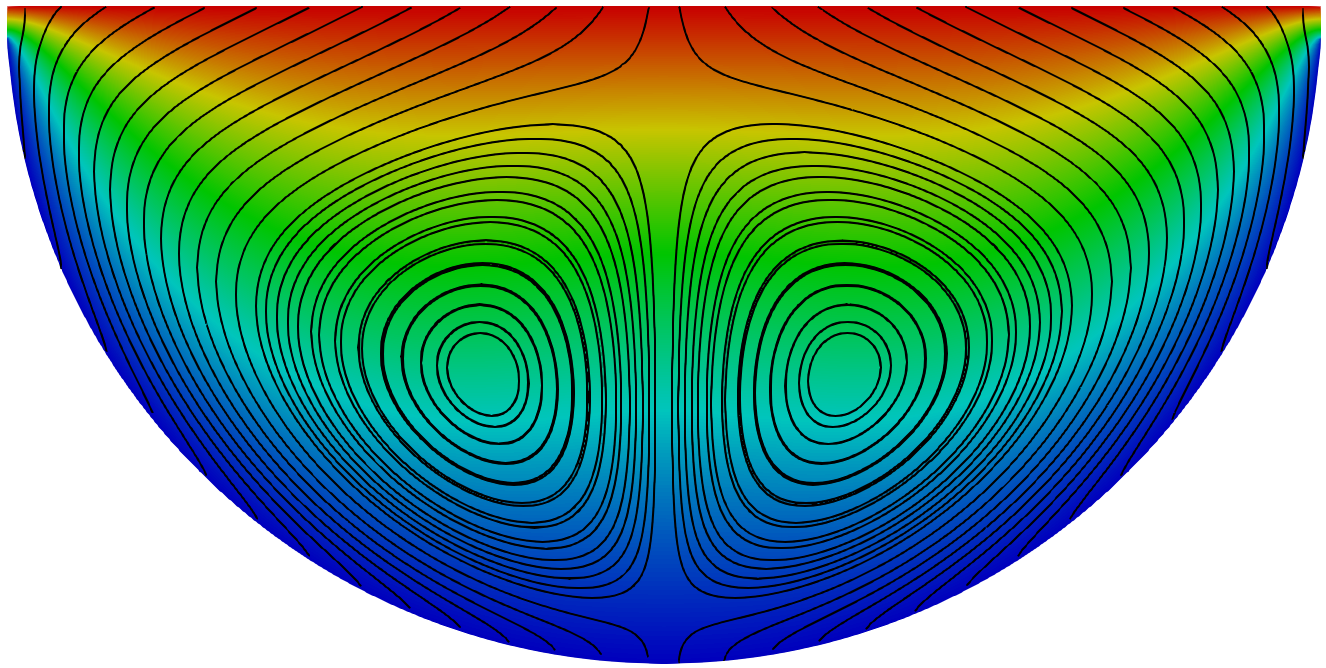}};
\node[anchor=north, inner sep=0, yshift=-30] (image) at (0,-3.8) {\includegraphics[width=0.44\textwidth]{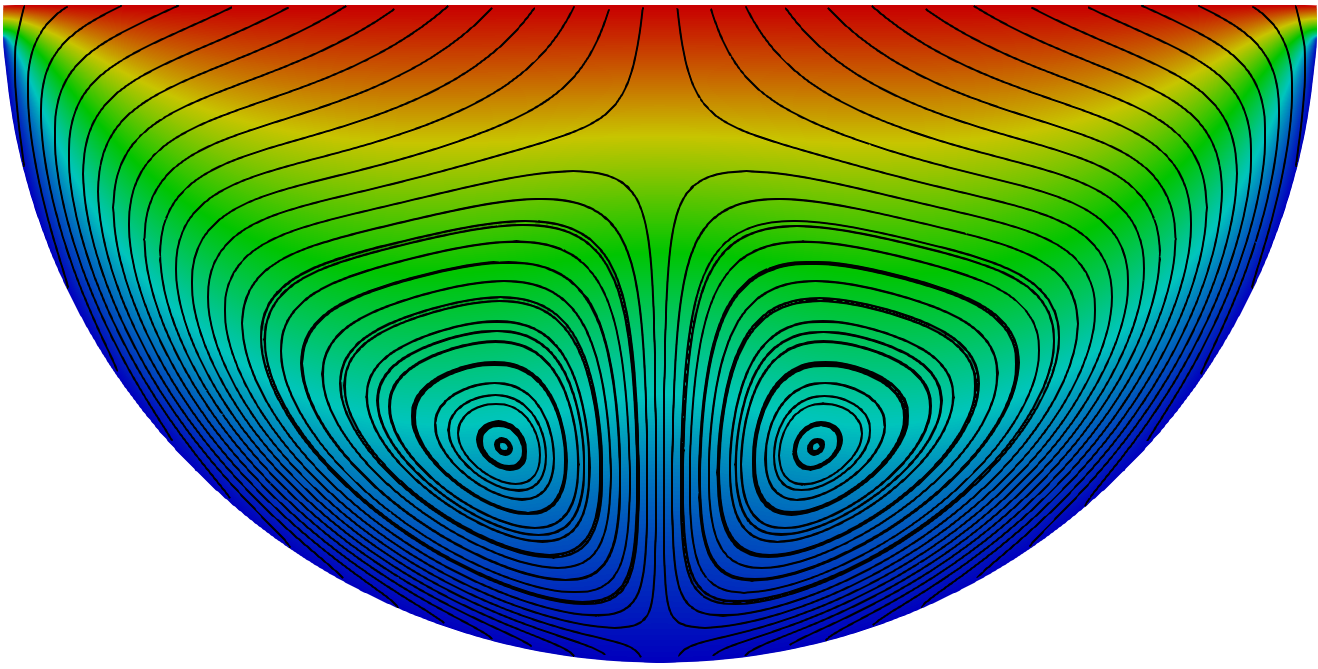}};
\node[anchor=north, inner sep=0,  xshift=100, yshift=-30] (image) at (6,-3.8) {\includegraphics[width=0.44\textwidth]{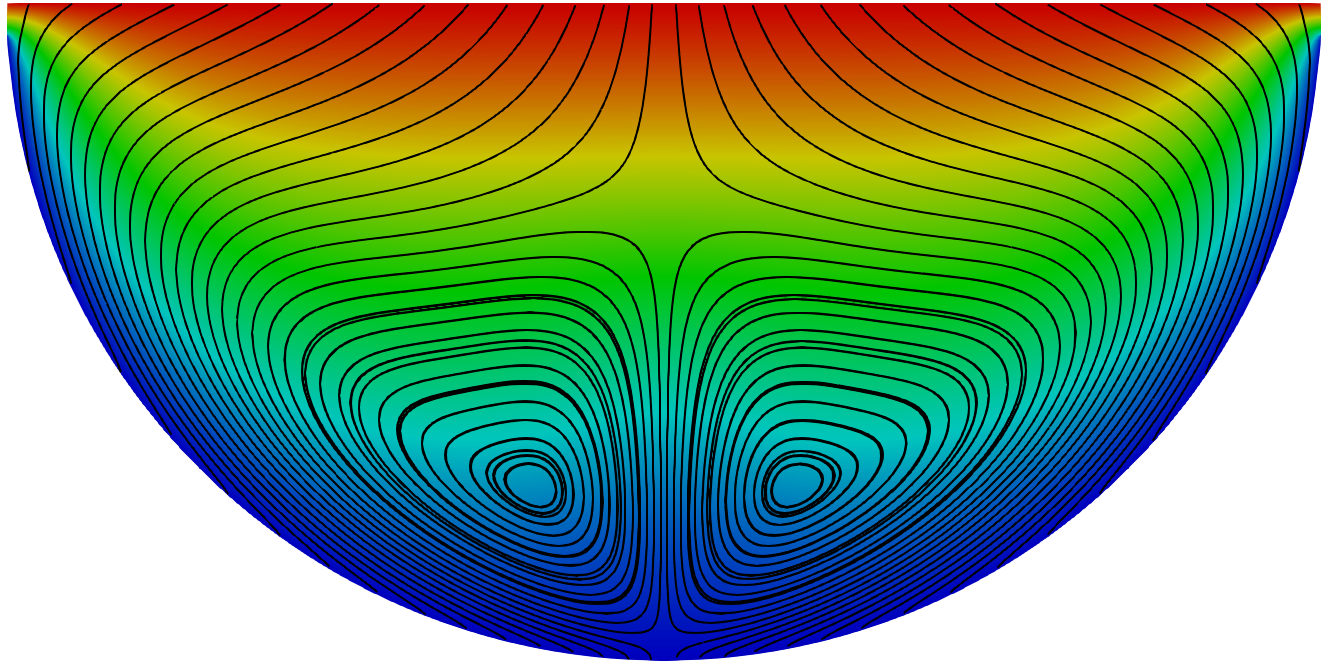}};
\node[anchor=south west,inner sep=0, xshift=40, yshift=-60] (image) at (2,-7) {\includegraphics[width=0.15\textwidth]{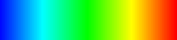}};
\node [below, xshift=40, yshift=-60] at (2,-7) { $0$};
\node [below, xshift=60, yshift=-60] at (4,-7) { $1$};
\node [above, xshift=50, yshift=-55] at (3,-6.5) { $T$};
\end{tikzpicture}
\end{center}
\vspace{-0.5\baselineskip}
\caption{Streamlines of the steady two-dimensional base flow and temperature field for the simulation cases (a) C1, (b) C2, (c) C3, and (d) C4.} 
\label{fig:baseflow}
\end{figure}
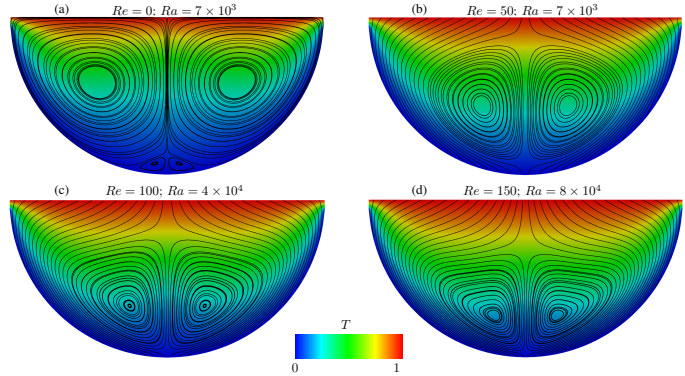

For the reminder of the paper, we shall proceed as follows to find and assess the actuation best suited to damp the growth of linear instabilities: 
First, the steady two-dimensional base flow solutions are obtained using direct numerical simulation (DNS) of (\ref{eq:u})-(\ref{eq:div_u}) together with the associated boundary conditions. Second, the linear stability analysis (LSA) of the two-dimensional base flow against three-dimensional perturbations is carried out by solving the eigenvalue problem for operator $\mathcal L$. These first two steps were previously carried out over an extensive range of governing parameters and wavenumbers $k$ in ~\cite{Kumar:JFM2020}. Here, we repeat the same approach for flows that are weakly supercritical so as to focus on cases where the instability is driven by a small number of unstable modes. The idea behind this strategy is that 
even in the fully nonlinear evolution, if the instability remains driven by a small enough number of modes, it may be enough to prevent the growth of the 
most unstable of them to stop the growth of instabilities altogether. This approach is not expected to be successful if the base flow is destabilised by a broad spectrum of fast growing unstable modes, as may be the case in more strongly supercritical cases.  On this basis, we select four typical weakly supercritical cases illustrating the different instability mechanisms identified in this previous work.
\begin{itemize}
\item C1 (${Re = 0; Ra=7 \times 10^3}$): This case corresponds to a purely convective base flow with zero inlet (and outlet)  mass flux, \emph{i.e.}, no fluid crosses the boundaries of the flow domain (see figure~\ref{fig:baseflow}(a)). The flow becomes unstable to a travelling wave at $Ra_c=5.975 \times 10^3$, through a supercritical Hopf bifurcation. The corresponding branch in the complex eigenmode spectrum is labelled type II. 
\item C2 (${Re = 50;  Ra=7 \times 10^3}$): This case features an inflow through the upper boundary as shown in figure~\ref{fig:baseflow}(b). The flow becomes unstable to a type II travelling wave through a supercritical Hopf bifurcation. 
\item C3 (${Re = 100;  Ra=4 \times 10^4}$): This case is similar to C2 but with more intense inflow. The base flow is presented in figure~\ref{fig:baseflow}(c).  At $Ra_c=3.621 \times 10^4$, it becomes unstable to a leading mode from a different branch (type I), that is still oscillatory but arises out of a subcritical Hopf bifurcation. 
\item C4 (${Re = 150; Ra=8 \times 10^4}$): Here the flow is mostly driven by the inflow (see figure~\ref{fig:baseflow}(d)). The instability corresponds to a further branch of the eigenvalue spectrum labelled type III and yields a non-oscillatory mode through a supercritical pitchfork bifurcation.
\end{itemize}  
Details of the above selected parameters are tabulated in table~\ref{tab:parameters}.

\setlength{\tabcolsep}{8pt}
\begin{table}
\begin{center}
\def~{\hphantom{0}}
\begin{tabular}{c c c c c  c}
Simulation \\ Case & $Re$ & $Ra$  & $(Ra/Ra_c)-1$ & $N_U$ & $k$\\[2 mm]
C1 & $0$   & $7 \times 10^3$ & $0.17$ & $2$ & $6$\\
C2 & $50$   & $7 \times 10^3$ & $0.13$ & $2$ & $6$ \\
C3 & $100$   & $4 \times 10^4$ & $0.10$ & $1$ & $4$\\
C4 & $150$   & $8 \times 10^4$ & $0.09$ & $1$ & $6$  
\end{tabular}
\caption{Parameters of the weakly supercritical cases, C1, C2, C3, and C4,  were selected for the analysis. Here $(Ra/Ra_c)-1$ represents the level of criticality, where $Ra_c$ is the critical Rayleigh number,  $N_U$ represents the number of unstable modes, and $k$ represents the most unstable mode. Note that $Ra_c$ for the each case is obtained from the table 2 of~\cite{Kumar:JFM2020}.}
\label{tab:parameters}
\end{center}
\end{table}

Additionally to this previous analysis, we now need to calculate adjoint modes to perform the receptivity and sensitivity analyses (RSA). These are obtained by solving the eigenvalue problem for the operator $\mathcal{L}^*$ using the same method as for the LSA. Third, the evolution of the normal mode targeted for suppression is calculated using the linearised Navier--Stokes equations (\ref{eq:direct}), to which a forcing term based on the 
adjoint eigenmode is added on the left hand side. {We use two types of actuation to that effect. As the unstable mode is suppressed by the forcing, its amplitude varies, and so does the amplitude of the actuation. Otherwise, once the unstable mode is suppressed, the actuation would act as an external force taking the flow away from its equilibrium. Hence, to assess the ability of an actuation based on the receptivity map to suppress the unstable mode, we use a forcing of the form:
\begin{equation}
{\bf f}(x,y,z,t) = \alpha(t)\Re\{A\hat{\bf q}^*(x,y)\exp{[ \sigma t + i(kz + \omega t + \phi)]}\},
\label{eq:forcing_adaptative}
\end{equation}
where $\alpha(t)$ represents the amplitude of the unstable mode normalised by its initial value. In practice, adapting the forcing to the amplitude in real time would require sensing and processing capable of extracting the evolution of the unstable mode instantaneously, which, in the harsh environment of continuous casting, is impractical. Instead, it is much easier to set a threshold on the sensor output above which an actuation of constant amplitude is applied. For this purpose, we use a simpler form of actuation:
\begin{equation}
{\bf f}(x,y,z,t) = \Re\{A\hat{\bf q}^*(x,y)\exp{[ \sigma t + i(kz + \omega t + \phi)]}\}.
\label{eq:forcing}
\end{equation}
In both cases, $A$ and $\phi$ represent the initial real amplitude and real phase of the forcing, respectively, relative to the unstable mode. Note that our strategy is not to choose an actuation intended to manipulate the eigenvalue associated with the leading eigenmode. This well-established technique entails applying a force whose linear dependence on  the unstable eigenmode shifts the leading eigenvalue of the linear evolution operator to the stable region. The topology of the optimal forcing map is, in this case, provided by the structural sensitivity, or base-flow sensitivity maps~\citep{marquet2008_jfm,GIANNETTI:JFM2007}. Instead, we chose to apply a predetermined force, \emph{i.e.}, that does not depend linearly on variable $\hat{\bf q}$. That actuation is designed to suppress the amplitude of the unstable mode from a small but finite initial value. Since the actuation does not linearly depend on $\hat{\bf q}$, the evolution equation is not strictly linear, its linear part does not change, and neither does the leading eigenvalue. The actuation {\bf f} has four components. The first three components represent mechanical forces, and the last component represents thermal forces. This step serves two purposes. First, it acts as a validation of the strategy. Since the forcing is derived from linearised adjoint equations, that ignore nonlinear interactions, it must at least be effective at damping the direct linear mode, to carry any hope of preventing the full nonlinear growth of the instability. Second, building the forcing on the topology of the adjoint mode involves a choice of amplitude and phase relative to the direct modes. To find out the optimal values of both, we carry out a series of linear simulations where they are varied.

Finally, with the knowledge of the optimal amplitude and phase of the actuation, we test the suppression of instability with its full nonlinear dynamics through 
three-dimensional (3D) DNS (the detail of individual simulations are given in \S~\ref{sec:Nonlinear_receptivity_forcing}).

\subsection{Numerical set-up}
The methodology outlined in the previous section involves four types of numerical computations: 2D (nonlinear) DNS, direct and adjoint eigenvalue problems, 3D evolution of individual eigenmodes through the direct linearised equations and 3D (nonlinear) DNS.
The solution of the direct and adjoint eigenvalue problems are found by means of the time-stepping method implemented and tested in detail in ~\cite{Kumar:JFM2020}. 
The novelty compared to this previous work is the solution of the adjoint eigenvalue problem, which was validated by making sure the eigenvalues obtained from both the LSA and RSA  yielded the same results down to machine precision.

\setlength{\tabcolsep}{20pt}
\begin{table}
\begin{center}
\def~{\hphantom{0}}
\begin{tabular}{c c c}
$N$  & $\sigma$   &   Relative error (\%) \\[2 mm]
$6$   & $0.21598$ & $1.67593$\\
$7$   & $0.21293$ & $0.24009$\\
$8$   & $0.21249$ & $0.03295$\\
$9$   & $0.21242$ & ---
\end{tabular}
\caption{We examine the relationship between the leading eigenvalues and the polynomial order $N$. The leading eigenvalues are computed on the mesh at $Re=150$, $Ra=8 \times 10^4$, and $k=6$. The relative error is calculated with respect to the case of the highest polynomial order ($N=9$).}
\label{tab:convergence_Re_150}
\end{center}
\end{table}

The non-linear governing equations  (\ref{eq:u})-(\ref{eq:div_u}), the direct perturbation equation (\ref{eq:direct}), and the adjoint equation (\ref{eq:adjoint}) are solved using the spectral-element code Nektar++~\citep{Cantwell:CPC2015,Moxey:CPC2020}. We adopted a spectral-element discretisation in the $x-y$ plane with a mesh consisting of 348 quadrilateral elements. For the three-dimensional simulations, we used a Fourier-based spectral method~\citep{Bolis:CPC2016} for discretisation in the ${\bf e}_z$ direction. The computational domain extends along ${\bf e}_z$ by $2\pi$. 
A third-order implicit-explicit (IMEX) method~\citep{Vos:IJCFD2011} is used for time-stepping. For all four types of numerical calculations, the time step was kept constant so that the maximum local Courant number $C_{\max}$ remained below unity everywhere in the domain at all times. Table~\ref{tab:Simulation_details} lists the values of time step $\Delta t$ for each set of simulations. 
The numerical implementation is described and tested in detail in ~\cite{Kumar:JFM2020}. 
Figure~\ref{fig:mesh_flow} shows the details of a two-dimensional $x-y$ mesh generated using the GMSH package~\citep{GeuzaineIJNME:2009} with polynomial order $N = 3$ as an example of spatial-spectral discretisation used in the $(x,y)$ plane. Elements at the edges are more densely packed than in the bulk, with a ratio of four between the edge sizes of the largest and the smallest elements. On each element, the flow variables are projected onto the polynomial basis represented at Gauss--Lobatto--Legendre points. 
As in our previous work, we perform a convergence test on the polynomial order for the leading eigenvalue for each case to ensure the solution is independent of the spectral order. For example, the leading eigenvalue computed for the simulation case C4 with the polynomial order $N=8$ differs by less than 0.04\% from the calculation performed with $N=9$. Convergence of the leading eigenvalue with the polynomial order is presented in table~\ref{tab:convergence_Re_150}. Similar convergence tests have been conducted for other simulation cases. The order of polynomials retained for each case is listed in table~\ref{tab:Simulation_details}. For all 3D DNS calculations, we have retained 32 Fourier modes along ${\bf e}_z$, which are deemed adequate based on our previous convergence test conducted in~\cite{Kumar:JFM2020}.

\section{Linear receptivity and sensitivity analyses}
\label{sec:receptivity_sensitivity_anal}
\begin{figure}
\begin{center}
\includegraphics[width=0.6\textwidth]{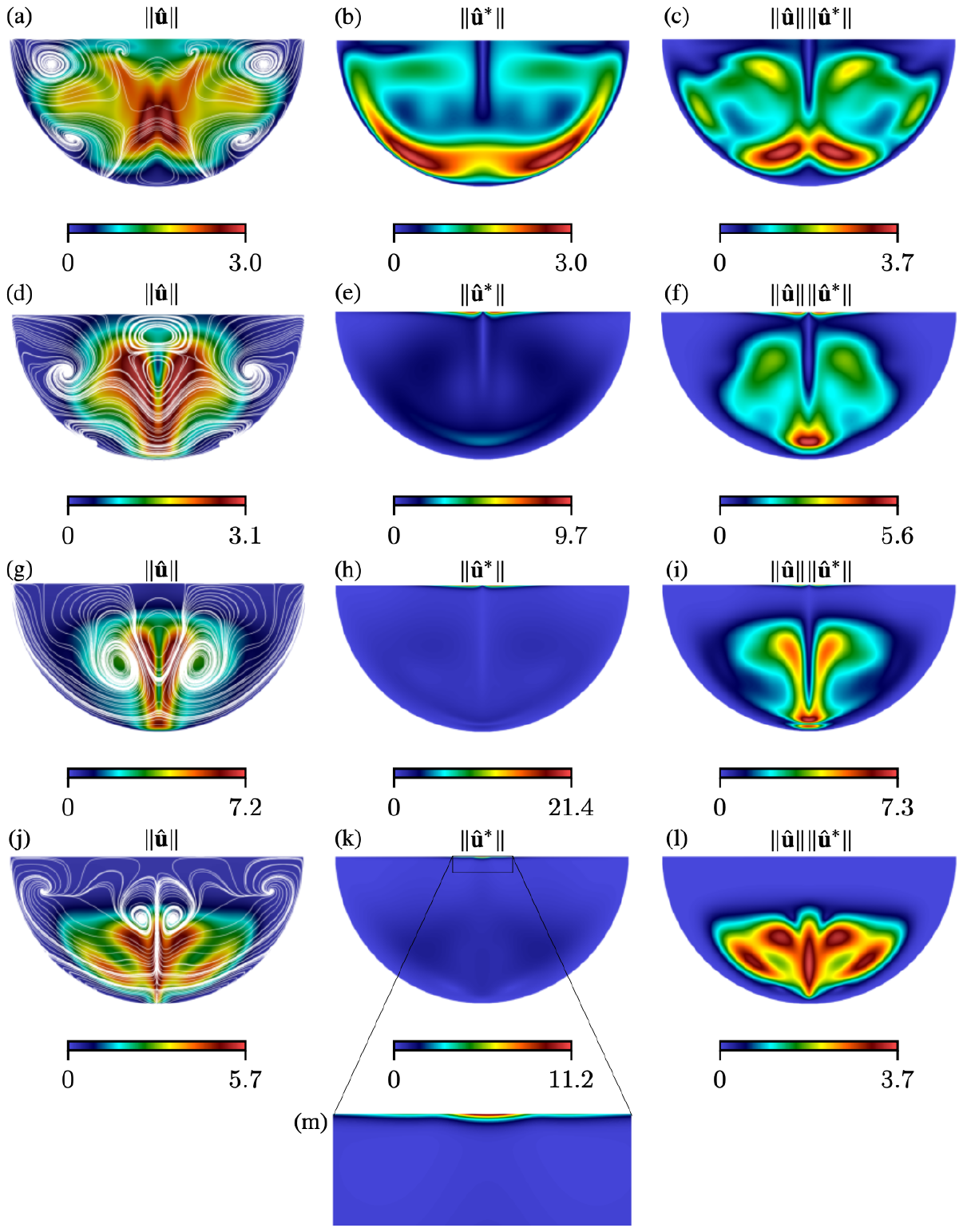}
\end{center}
\caption{Spatial distribution of the velocity field modulus ($\|\hat{\bf u}\|$), receptivity to momentum forcing  ($\|\hat{\bf u}^*\|$), and the Frobenius norm of the momentum structural sensitivity ($\|\hat{\bf u}\|\|\hat{\bf u}^*\|$) for the simulation cases: (a)-(c) C1;  (d)-(f) C2; (g)-(i) C3; (j)-(l) C4. (m) Represents the zoomed region of $\|\hat{\bf u}^*\|$ for C4. The streamlines in (a), (d), (g), and (j) represent the real part of the unstable eigenmode ($\Re({\hat{u}}){\bf e}_x + \Re({\hat{v}}){\bf e}_y$) in the $x-y$ plane. } 
\label{fig:receptivity_sensitivity}
\end{figure}
The aim of this section is to identify the most effective location within the flow domain for an actuation to suppress instabilities using RSA. The receptivity analysis returns a map of the physical locations within the flow domain where one can act on the instabilities by applying an external actuation. On the other hand, the sensitivity analysis identifies the location in the flow domain where this external actuation would incur the greatest flow alteration. For this purpose, we shall discuss each of the cases C1-C4 outlined in \S~\ref{ssec:Methodology_parameters}. 

For C1, the base flow, represented on figure~\ref{fig:baseflow}(a), consists of two symmetric baroclinically-driven recirculation cells driven along the solidification front, from the maximum baroclinicity regions in the upper left and right corners. Both cells meet on the axis near the bottom of the domain and drive a strong upward jet there. The leading unstable mode of type II (with critical parameters $k_c=6.3$, $Ra_c = 5.975 \times 10^3$) arises out of shear instability near the location of the maximum velocity along that jet (see figure~\ref{fig:receptivity_sensitivity}(a)). It consists of a wave travelling in the $z$ direction, and appears through a supercritical Hopf bifurcation. 

The velocity field modulus associated to the adjoint mode $\|\hat{\bf u}^*\|$, which represents receptivity, is displayed in figure~\ref{fig:receptivity_sensitivity}(b). The most receptive region is located at the wall, where the magnitude of two jets is strongest.  The product of direct and adjoint modes represents the sensitivity and is plotted in figure~\ref{fig:receptivity_sensitivity}(c).  This plot shows that the sensitivity is symmetric about the central axis, and strong towards the bottom of the cavity. 

For C2, the presence of the inflow through the top boundary opposes the upward jet seen in C1 and therefore suppresses the baroclinically-driven recirculation. These are displaced downwards as a result and the shear is less concentrated near the symmetry axis as represented in figure~\ref{fig:baseflow}(b). Accordingly, the unstable mode ($k_c=6.2$, $Ra_c=6.168 \times 10^3$, figure~\ref{fig:receptivity_sensitivity}(d)) is more widely spread along the two directions of space than in C1 and stretches down to the bottom of the cavity. It also separates into two lobes corresponding to each recirculation, whilst keeping a maximum intensity where they meet at the bottom of the cell. The most receptive region (see figure~\ref{fig:receptivity_sensitivity}(e)) lies just below the upper surface, where the fluid flows into the cavity from either side of the central axis. The sensitivity, as shown in figure~\ref{fig:receptivity_sensitivity}(f), is particularly strong at the bottom of the domain, where oscillations are caused by the instability. The topology of the receptive mode has two consequences: First, the most effective location to act on the unstable mode is not in the bulk of the flow, but near the top surface. Since the inflow directly impacts this region, altering the inflow profile may offer an effective means of applying optimal actuation. Second, if instead of attempting to control the unstable mode, one follows a control strategy consisting in modifying the base flow to stabilise the unstable mode, the actuation is best applied at the bottom of the cavity. This illustrates that the two approaches involve very different types of actuation. For the purpose of the application to casting, the bottom of the cavity corresponds to the locus of the solidification front, which is the least accessible. The inlet, by contrast, is much easier to access through the upper free surface.

In case C3, the greater inflow compared to C2 further suppresses the convective cells. These now become confined to the lower half of the domain, and extend over less then a third of its width (see figure~\ref{fig:baseflow}(c)). The topology of the unstable mode ($k_c=4.1$, $Ra_c=3.621 \times 10^4$, figure~\ref{fig:receptivity_sensitivity}(g)) remains similar to that of C2, but the lobes of greater energy are now separated along the jet to join only near the bottom where the mode's energy is still maximum. The receptivity is again concentrated near the top surface but has further contracted in size (see figure~\ref{fig:receptivity_sensitivity}(h)). The topology of the sensitivity map still shows a maximum in the lower part of the sump, albeit more extended  towards its centre, along the main central jet. As in C2, the regions of actuation for controlling the unstable mode and for structural sensitivity differ, with the former located in a much more accessible region of the flow in the context of continuous casting.

Lastly, case C4 corresponds to a different regime where the inflow further suppresses the convective cells (see figure~\ref{fig:baseflow}(d)). The unstable mode  ($k_c=6.0$, $Ra_c=7.345 \times 10^4$, figure~\ref{fig:receptivity_sensitivity}(j)) adopts a different topology with a sharp maximum along the central axis and four lobes extending either side of it in the middle of the bulk and near the bottom wall. These correspond to the location where streamlines are at maximum angle with the vertical direction. Somewhat surprisingly, despite a very different topology in their leading eigenmode, the receptivity modes in C3 and C4  exhibit relatively similar topologies, both of them being sharply concentrated in the middle of the free surface. The receptive mode of C4, however, is concentrated over an even smaller region (see figure~\ref{fig:receptivity_sensitivity}(k)). Additionally, in contrast to C2 and C3, the receptivity mode for C4 features a maximum exactly in the middle of the inlet, visible on magnified figure \ref{fig:receptivity_sensitivity}(m) whereas the receptivity modes for C2 and C3 are split into two lobes, each with a point of maximum intensity either side of the centre of the free surface. The sensitivity map occupies practically the entire lower half of the domain and, as such, remains difficult to access unlike the region highlighted by the receptivity map.

To summarise, despite different instability mechanisms, cases C2, C3, and C4 all exhibit receptivity maps showing strong localisation near the surface, whereas their sensitivity maps show localisation in the lower part of the domain. Since the lower half is practically inaccessible in the casting process, the classical strategy of altering the base flow to stabilise, which would require aligning the forcing with the sensitivity map, is not practical. By contrast, attempting to actively control the unstable mode requires a forcing aligned with the receptivity map, which is conveniently located near the upper surface.  Additionally, since the receptivity maps in cases C2, C3, and C4, have practically no overlap with the sensitivity maps, applying a forcing based on the former will likely not affect the structural stability of the problem. In this sense, the influence of the forcing on the base flow has little effect. 
The receptivity and sensitivity maps for case C1 are less localised but still exhibit maxima at different locations: on the side of the lower boundary for the receptivity map and closer to the main jet and further inside the bulk for the sensitivity map. In practice, this makes both strategies equally difficult to implement. Analysis of the components contributing to the sensitivity map provides insight into the feedback mechanism underpinning the growth of the unstable mode \citep{GIANNETTI:JFM2007}. This is systematically investigated through the sensitivity tensor provided in Appendix \ref{sec:sensitivity_tensor}.
\section{Linear response to an actuation based on receptivity modes}
\label{sec:Linear_receptivity_forcing}
Having now identified the topology of the receptivity modes and their likely effect on the instability, we shall now proceed to use these modes as a basis for the design of an actuation. The main question is whether applying such receptivity-based actuation during the period where instability would grow indeed alters the growth of the instability.  With the application to the casting of alloys in mind, we would indeed seek to at least stifle, and possibly prevent the growth of the instability as much as possible. The mathematical expression of the corresponding actuation is provided by equation (\ref{eq:forcing}). Importantly, while the first three components of ${\bf f}(x,y,z,t)$ indeed represent a force density applied in the three components of the Navier--Stokes equation, the fourth component applies to the energy equation (\ref{eq:T}) and, therefore, represents a heat source. The full actuation, therefore, comprises both a momentum and an energy source in the general case. 

\subsection{Methodology}
In the first instance, we seek the linear response of the system. This step acts as a validation, as the sensitivity analysis already provided us with an indication of the linear response we should expect. Two very important aspects still remain to be clarified by calculating the linear evolution of the leading eigenmode under the effect of the actuation: first, the timescale of the response and the duration of the effect are not considered in the receptivity analysis. Second, the receptivity does not specify  how the relative amplitude and phase of the receptive mode affect the linear response.  Indeed, in equations~(\ref{eq:forcing_adaptative}) and (\ref{eq:forcing}), the amplitude and phase of the actuation are both relative to the leading eigenmode. Since the receptivity mode is the solution of a 
homogeneous linear problem, none of them is specified. On the other hand the relative amplitude and phase of the actuation can be expected to greatly influence the system's response to it. For these two reasons, and to identify the combination of relative phase and amplitude that optimises the suppression of the instability, we run a series of linear simulations by adding a source term representing the receptivity-based actuation on the right-hand side of (\ref{eq:direct}):
\begin{equation}
\frac{\partial {\bf q}^\prime}{\partial t}  = \mathcal{L}  {\bf q}^\prime + {\bf f}.
\end{equation}
We conduct a parametric study for the actuation of constant amplitude (\ref{eq:forcing}), varying the phase $\phi$ from $0$ to $2\pi$ in increments of $\pi/10$ and the amplitude between 0.01 and 0.5 for most of the cases. Then, we conduct a single linear simulation with adaptive actuation amplitude (\ref{eq:forcing_adaptative}), using the most effective combination of $A$ and $\phi$ found in the parametric analysis, to assess whether the receptivity-based actuation indeed fully damps the unstable mode asymptotically.
\begin{figure}
\begin{center}
\includegraphics[width=0.9\textwidth]{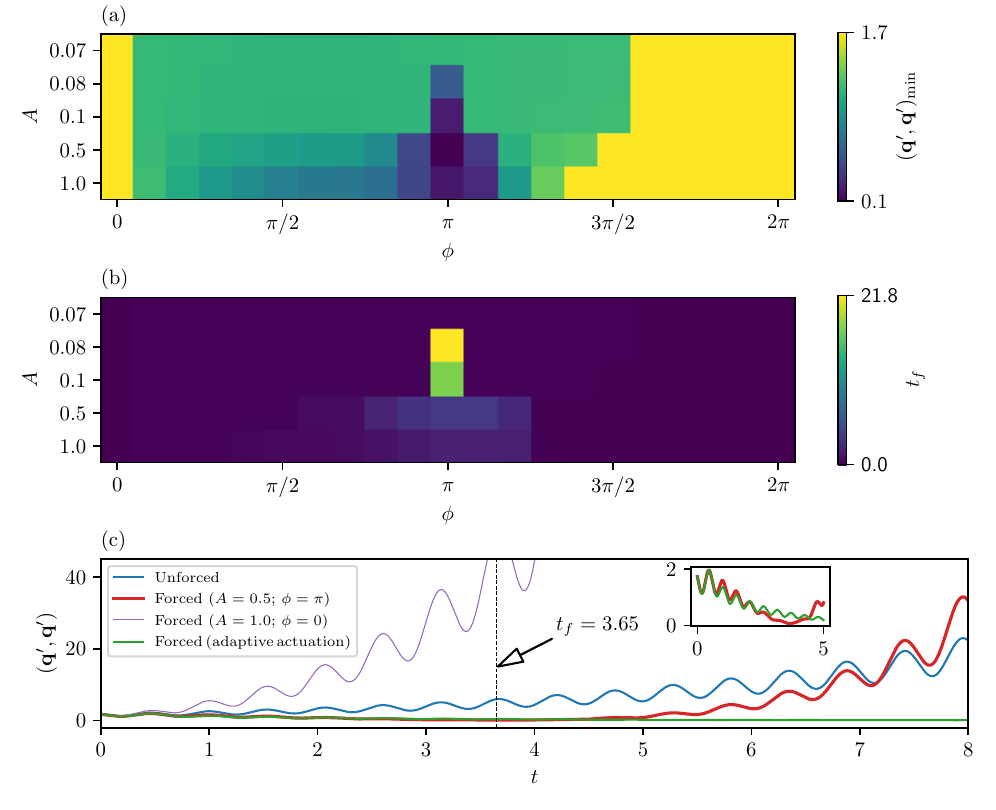}
\end{center}
\vspace{-\baselineskip}
\caption{For  the simulation case C1 ($Re=0$, $Ra=7 \times 10^3$, and $k=6$): (a) Minimum value of the strength of perturbation $({\bf q}^\prime, {\bf q}^\prime)_{\min}$ as a function of the  amplitude $A$ and the phase $\phi$ of the linear receptivity-based actuation. (b) Dependence of the time $t_f$ at which $({\bf q}^\prime, {\bf q}^\prime)_{\min}$ occurs on the amplitude $A$ and phase $\phi$. (c) The strength of perturbation $({\bf q}^\prime, {\bf q}^\prime)$ as a function of time $t$ for both unforced and forced cases. The inset represents the evolution of both constant amplitude actuation and adaptive actuation from $t=0$ to $t \simeq t_f$.}  
\label{fig:Linear_Receptivity_Forcing_Re_0}
\end{figure}
Each linear simulation is initiated using the unstable mode obtained from the LSA as the initial condition with amplitude such that the normalisation condition (\ref{eq:normalisation_condition}) is satisfied, from which both $A$ and $\phi$ are fixed. In the linear simulation, normal modes are decoupled from each other so the time-dependent solution obtained by initialising the solution with a single normal mode reflects the evolution of that particular mode only. As such, linear simulations provide a direct measure of the ability of the actuation based on the receptivity mode to affect the evolution of the unstable mode. To quantitatively asses this effect, we monitor the time-dependent energy of the mode,  $({\bf q}^\prime, {\bf q}^\prime)$. The lowest value $({\bf q}^\prime, {\bf q}^\prime)_{\min}$ reached by this quantity  gives an indication of the damping achieved by the actuation, and the time  of occurrence of this minimum measures the timescale over which this damping is  achieved, denoted as $t_f$. 
Going back to the example of the continuous casting of alloys as one of the motivations for this work, the unstable mode may not need to be suppressed indefinitely. Instead, maintaining the unstable mode to a low level for the entire duration of the casting operation, which is finite, would be sufficient to ensure it does not impact the final quality of the alloy. Hence the importance of $t_f$ is both fundamental and practical. 

\subsection{Analysis of cases C1-C4}
We shall now examine the outcome of this approach in each of the C1-C4 cases, defined in \S~\ref{ssec:Methodology_parameters}. Figure \ref{fig:Linear_Receptivity_Forcing_Re_0}(c)  illustrates two examples of the evolution of $({\bf q}^\prime, {\bf q}^\prime)$ for C1. The blue curve represents a simulation with no actuation, \emph{i.e.}, one with $A=0$ amplitude. As predicted by LSA, the instability grows exponentially with oscillations of frequency determined by the imaginary part of the mode's eigenvalue. The red curve represents the case with receptivity-based actuation of amplitude $A=0.5$  and phase $\phi=\pi$. In this case, $({\bf q}^\prime, {\bf q}^\prime)$ decreases over time and reaches its minimum value at $t_f=3.65$. For $t>t_f$, $({\bf q}^\prime, {\bf q}^\prime)$ increases, and crosses the blue curve at $t=7.16$. Hence, in this particular case,  the actuation results in an effective suppression of the instability until $t=t_f$. 
Now, to determine the optimal value of  $A$ and $\phi$, \emph{i.e.}, values that achieve the smallest value of $({\bf q}^\prime, {\bf q}^\prime)_{\min}$, 
we vary the phase in the range of $0$ to $2\pi$, and for amplitudes between 0.07 to 1.0. The grid map of the corresponding values of $({\bf q}^\prime, {\bf q}^\prime)_{\min}$ is shown in figure \ref{fig:Linear_Receptivity_Forcing_Re_0}(a) for C1. It turns out that $A=0.5$  and $\phi=\pi$ are the optimal values of amplitude and phase with$({\bf q}^\prime, {\bf q}^\prime)_{\min} = 0.06$. Simulation conducted with adaptive actuation (\ref{eq:forcing_adaptative}) shows that the unstable mode decays very similarly to the case of constant amplitude actuation until $t\simeq t_f$, but continues to decay asymptotically to zero as $t \to \infty$. This shows that the receptivity-based actuation indeed suppresses the unstable mode completely.
\begin{figure}
\begin{center}
\includegraphics[width=0.9\textwidth]{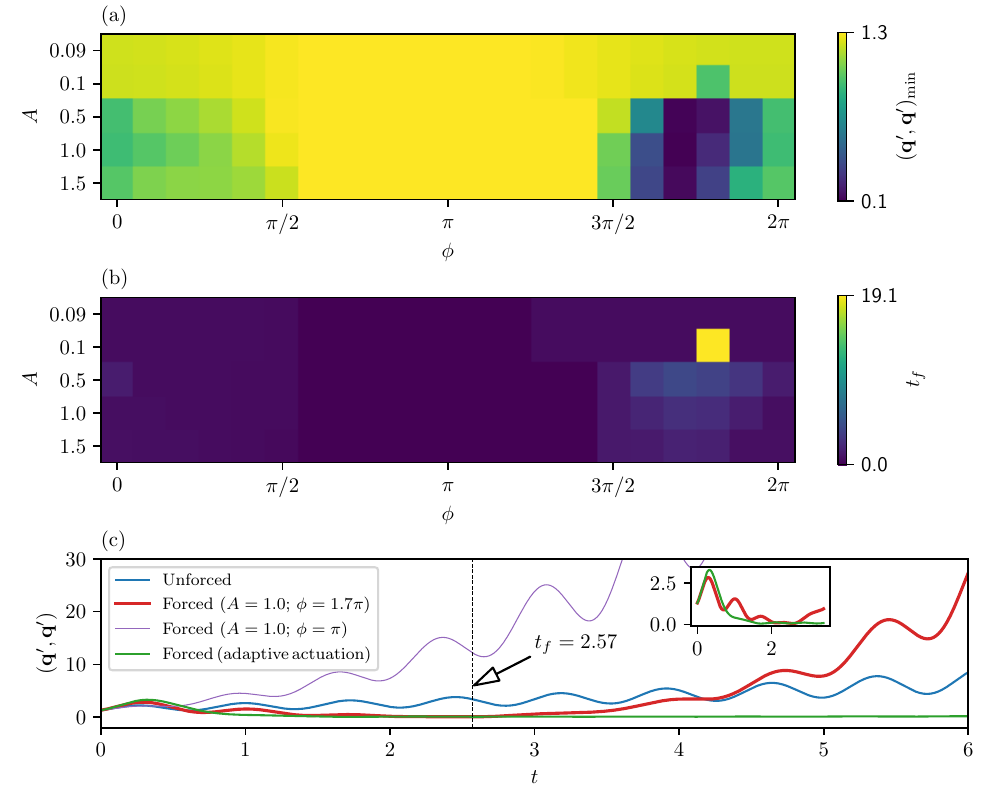}
\end{center}
\vspace{-\baselineskip}
\caption{For  the simulation case C2 ($Re=50$, $Ra=7 \times 10^3$, and $k=6$): (a) Minimum value of the strength of perturbation $({\bf q}^\prime, {\bf q}^\prime)_{\min}$ as a function of the  amplitude $A$ and the phase $\phi$ of the linear receptivity-based actuation. (b) Dependence of the time $t_f$ at which $({\bf q}^\prime, {\bf q}^\prime)_{\min}$ occurs on the amplitude $A$ and phase $\phi$. (c) The strength of perturbation $({\bf q}^\prime, {\bf q}^\prime)$ as a function of time $t$ for both unforced and forced cases. The inset represents the evolution of both constant amplitude actuation and adaptive actuation from $t=0$ to $t \simeq t_f$.} 
\label{fig:Linear_Receptivity_Forcing_Re_50}
\end{figure}
\begin{figure}
\begin{center}
\includegraphics[width=0.9\textwidth]{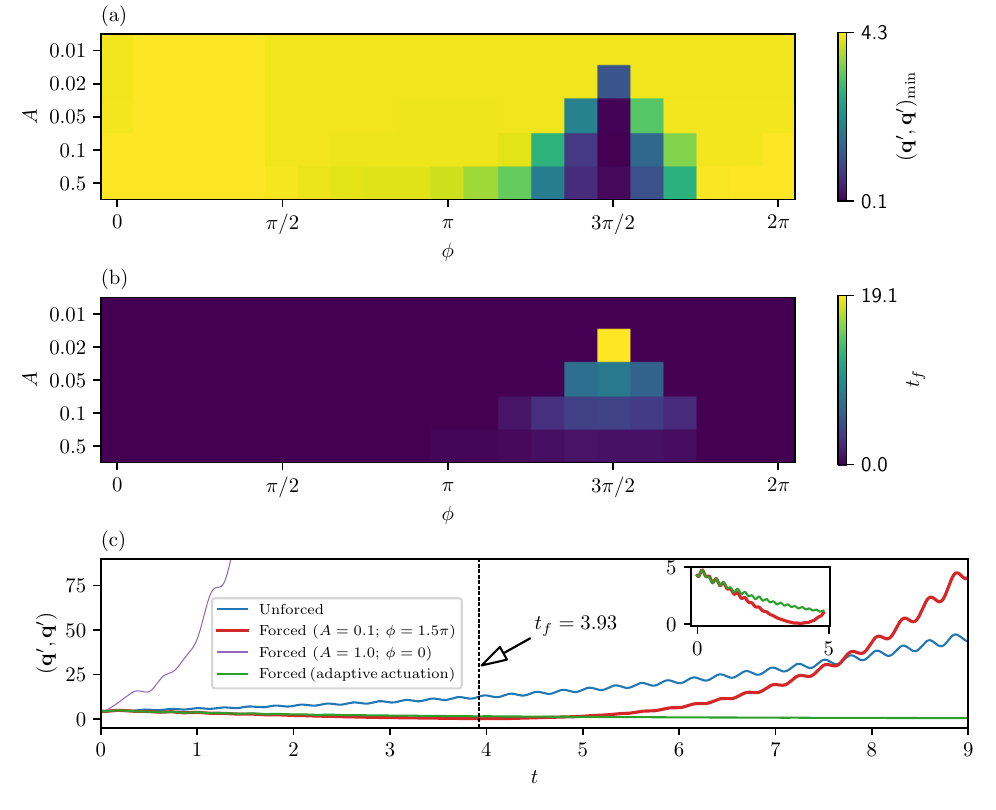}
\end{center}
\vspace{-\baselineskip}
\caption{For  the simulation case C3 ($Re=100$, $Ra=4 \times 10^4$, and $k=4$): (a) Minimum value of the strength of perturbation $({\bf q}^\prime, {\bf q}^\prime)_{\min}$ as a function of the  amplitude $A$ and the phase $\phi$ of the linear receptivity-based actuation. (b) Dependence of the time $t_f$ at which $({\bf q}^\prime, {\bf q}^\prime)_{\min}$ occurs on the amplitude $A$ and phase $\phi$. (c) The strength of perturbation $({\bf q}^\prime, {\bf q}^\prime)$ as a function of time $t$ for both unforced and forced cases. The inset represents the evolution of both constant amplitude actuation and adaptive actuation from $t=0$ to $t \simeq t_f$.} 
\label{fig:Linear_Receptivity_Forcing_Re_100}
\end{figure}

Further calculations were performed for C2 and C3, which, despite corresponding to different instability branches, return similar results, presented in figures~\ref{fig:Linear_Receptivity_Forcing_Re_50}  and~\ref{fig:Linear_Receptivity_Forcing_Re_100}. As compared to C1, C2 has an optimal amplitude that is twice as large, with a phase shift of $1.7\pi$ radians. Therefore, the optimal values are  $A=1.0$  and  $\phi=1.7\pi$. The value of $t_f$ is slightly reduced to 2.57. On the other hand, for C3, the optimal amplitude is reduced to 0.1, and the optimal phase is shifted to $1.5\pi$. The value of $t_f$, however, is slightly increased to 3.93, which implies that the instability can be attenuated for a slightly longer period of time. As in C1, the adaptive actuation (\ref{eq:forcing_adaptative}) leads to a decay similar to the constant amplitude actuation until $t\simeq t_f$, but it continues to decay asymptotically to zero as $t \to \infty$, so in C2 and C3 too, the receptivity-based actuation fully suppresses the unstable mode.
\begin{figure}
\begin{center}
\includegraphics[width=0.9\textwidth]{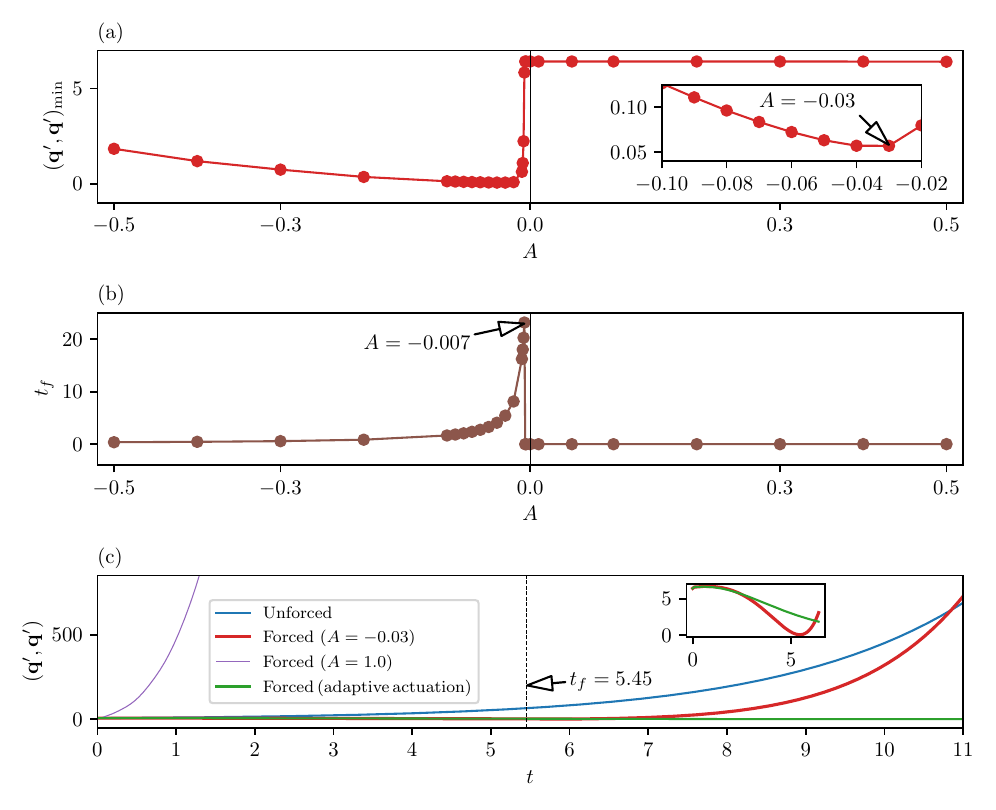}
\end{center}
\vspace{-\baselineskip}
\caption{For  the simulation case C4 ($Re=150$, $Ra=8 \times 10^4$, and $k=6$):(a) Minimum value of the strength of perturbation $({\bf q}^\prime, {\bf q}^\prime)_{\min}$ as a function of the  amplitude $A$ of the linear receptivity-based actuation. Inset represents the amplitude $A=-0.03$ which corresponds to the lowest value of $({\bf q}^\prime, {\bf q}^\prime)_{\min}$. (b) Dependence of the time $t_f$ at which $({\bf q}^\prime, {\bf q}^\prime)_{\min}$ occurs on the amplitude $A$. (c) The strength of perturbation $({\bf q}^\prime, {\bf q}^\prime)$ as a function of time $t$ for both unforced and forced cases. The inset represents the evolution of both constant amplitude actuation and adaptive actuation from $t=0$ to $t \simeq t_f$.} 
\label{fig:Linear_Receptivity_Forcing_Re_150}
\end{figure}

C4 differs from the other cases in that the leading eigenmode and its adjoint, which represents receptivity, are non-oscillatory. Consequently, the actuation is also non-oscillatory and we do not need to determine the optimal phase, only the optimal amplitude. In the absence of a phase, the possibility still remains to reverse the actuation, which we incorporated into the sign of $A$.  The variations of $({\bf q}^\prime, {\bf q}^\prime)_{\min}$ with $A$ are shown in figure \ref{fig:Linear_Receptivity_Forcing_Re_150}(a). For $A\geq0$, $({\bf q}^\prime, {\bf q}^\prime)_{\min}$ remains very close to the initial value of $({\bf q}^\prime, {\bf q}^\prime)$ so no suppression is achieved. For $A<0$, by contrast, we observe a sudden drop in the value of $({\bf q}^\prime, {\bf q}^\prime)_{\min}$ at very small amplitudes, with a minimum value at $A=-0.03$ (see inset of figure \ref{fig:Linear_Receptivity_Forcing_Re_150}(a)).

Figure \ref{fig:Linear_Receptivity_Forcing_Re_150}(c) shows the evolution of $({\bf q}^\prime, {\bf q}^\prime)$  over time for both the unforced and forced cases with $A=-0.03$. It appears that the actuation achieves a suppression of the non-oscillatory mode up to time $5.45$. Here again, the adaptive actuation (\ref{eq:forcing_adaptative})  suppresses the unstable mode completely and acts over the same timescale $t_f$ as the constant amplitude actuation.
  \setlength{\tabcolsep}{20pt}
\begin{table}
\begin{center}
\def~{\hphantom{0}}
\begin{tabular}{c | c c| c c|}
 \multicolumn{5}{c}{\hspace{2.3 cm} Suppression-optimising \hspace{1.9cm} Time-optimising} \\[1 mm]
 \hline
Simulation Case & $\sigma t_f$  & $\frac{({\bf q}^\prime, {\bf q}^\prime)_{\min}}{({\bf q}^\prime, {\bf q}^\prime)_{t=0}}$ & $\sigma t_f$  & $\frac{({\bf q}^\prime, {\bf q}^\prime)_{\min}}{({\bf q}^\prime, {\bf q}^\prime)_{t=0}}$ \\[2 mm]
C1   & $0.57$  & $0.037$ & 3.43 & 0.314\\
C2   & $0.31$  & $0.040$ & 2.34 & 0.733\\
C3   & $0.51$  & $0.013$ & 2.51 & 0.272\\
C4   & $1.16$  & $0.009$ & 4.93 & 0.910
\end{tabular}
\caption{The normalised value of $t_f$ with respect to the growth rate $\sigma$ for various suppression-optimising and time-optimising simulation cases. Comparison of $({\bf q}^\prime, {\bf q}^\prime)_{\min}$ relative to the initial value of $({\bf q}^\prime, {\bf q}^\prime)$ is shown for each suppression-optimising and time-optimising simulation cases.}
\label{tab:Simulation_details_LSA}
\end{center}
\end{table}

\subsection{The salient features of the linear response to an actuation of constant amplitude}
In all cases, the receptivity-based actuation with adaptive amplitude suppresses the unstable mode asymptotically. This result shows that the theoretical foundations for the receptivity-based instability control are sound. However, the actuation with constant amplitude is more useful in practical situations, but induces a more complex response, which we now discuss in more detail. All four cases show strong similarities, despite the differences in the nature of the unstable modes.  First, the linear actuation has a stabilising effect up to a finite time $t_f$, effectively suppressing the instability. Beyond this point, the unstable mode grows again and at a faster pace than when unforced. This means that for $t>t_f$, the linear actuation becomes \emph{destabilising}. The reason is that, since the base flow is fixed in the linear equations we simulate, the actuation's only effect is to suppress the unstable mode. Once the mode is suppressed, the system is back to the equilibrium point defined by the steady base flow. The actuation then acts as an additional force, moving the system away from this equilibrium. Nonlinear simulations are required to verify whether the base flow is indeed affected by the actuation and whether the phenomenology observed in the linear framework is indeed valid. Table \ref{tab:Simulation_details_LSA} shows that $t_f$ remains of the order of the inverse of the growth rate of the unstable mode, as the ratio $\sigma t_f$ remains of the order of unity for all suppression-optimising cases. The term ``suppression-optimising'' refers to seeking the amplitude and phase of the forcing that achieve the maximum reduction in mode amplitude. The reduction in mode amplitude achieved at $t=t_f$ is of approximatively two orders of magnitudes (see table \ref{tab:Simulation_details_LSA}) for the suppression-optimising cases. This suggests that the actuation is more effective when applied for a finite time $t_f$, then stopped until the unstable mode regrows to its initial amplitude (\emph{i.e.} during a time $\sim\sigma^{-1}$), at which point the procedure can be repeated. Whether this strategy is realistic depends on how this pattern is affected by nonlinearities.

This pattern also raises the question whether instead of applying suppression-optimising, it may be beneficial to seek configurations that achieve the largest value of $t_f$ so as to maximise the time during which the suppression is effective. The term ``time-optimising"  is introduced to denote the search for amplitude and phase settings that maximise the value of $t_f$. The corresponding optimal values are shown on figures \ref{fig:Linear_Receptivity_Forcing_Re_0}(b), \ref{fig:Linear_Receptivity_Forcing_Re_50}(b), \ref{fig:Linear_Receptivity_Forcing_Re_100}(b), and \ref{fig:Linear_Receptivity_Forcing_Re_150}(b). Unsurprisingly, actuation parameters optimising the suppression and $t_f$ differ. We shall compare how both optimals perform on the nonlinear response in the next section.

In addition to identifying optimal suppression parameters, our results highlight that certain combinations of amplitude and phase can instead enhance instability. For instance, as shown in figure \ref{fig:Linear_Receptivity_Forcing_Re_0}(c), the case $A=1$ and $\phi=0$ leads to the most destabilising effect for $Re=0$, and similarly, as shown in figure \ref{fig:Linear_Receptivity_Forcing_Re_100}(c), for $ Re=100 $. In contrast, for $Re=50$ (figure \ref{fig:Linear_Receptivity_Forcing_Re_50}(c)), the configuration $A=1$ and  $\phi=\pi$ is more destabilising than $ A=1 $ and $ \phi=0 $.  This observation underscores the importance of carefully selecting both the amplitude and phase in actuation design, as an inappropriate choice can inadvertently amplify perturbations rather than suppressing them.

\section{Nonlinear response to an actuation based on receptivity} \label{sec:Nonlinear_receptivity_forcing}
\subsection{Methodology}
\setlength{\tabcolsep}{6pt}
\begin{table}
\begin{center}
\def~{\hphantom{0}}
\begin{tabular}{c c c c c c  c c c c}
	Case &$Re$  & $Ra$  & $N$ & $\Delta t$& $\mathcal{R}_A$ &   $t_f$  & $t_{\rm eff}$  & $\frac{t_{\rm eff}}{t_f}$ & $\eta_a$\\[2 mm]
C1& $0$   & $7 \times 10^3$ & $6$ & $5 \times 10^{-3}$ & $1 \times 10^{-2}$ & $3.65$ & 19.1 & 5.23 & 5.21\\
C2& $50$   & $7 \times 10^3$ & $7$ & $1 \times 10^{-4}$ &  $4 \times 10^{-4}$ & $2.57$ & 17.8 & 6.93& 6.30\\
C3& $100$   & $4 \times 10^4$ &$7$ & $1 \times 10^{-4}$ & $1 \times 10^{-3}$ & $3.92$ & 32.5 & 8.29 & 8.27\\
C4 & $150$   & $8 \times 10^4$ &  $8$ & $5 \times 10^{-5}$ & $4 \times 10^{-5}$ & $5.45$ & 21.4 & 3.93 & 3.93 
\end{tabular}
	\caption{Parameters of our numerical calculations: Reynolds number $Re$, Rayleigh number $Ra$, order of polynomial $N$, time step $\Delta t$. Amplitude ratio $\mathcal{R}_A$, the forcing cut-off time $t_f$, nonlinear efficiency time $t_{\rm eff}=t_2-t_1$, and actuation efficiency $\eta_a=t_R/t_f$ based on recovery time $t_R$ such that $({\bf q}^\prime, {\bf q}^\prime)_{t=t_R}=({\bf q}^\prime, {\bf q}^\prime)_{t=0}$ for the 3D DNS.}
\label{tab:Simulation_details}
\end{center}
\end{table}
The linear simulations confirmed that an actuation designed from the topology of the receptivity, with a frequency matched to the leading eigenmode was indeed capable of stifling the 
linear mechanism responsible for the growth of the leading eigenmode. We also identified the relative amplitude and phase that maximised its suppression. In reality, the finite amplitude of the perturbation, whether subject to the actuation or not, activates a nonlinear response of the system. Although nonlinearities may not measurably deflect the perturbation's evolution in its  early stages, when its amplitude is still small, it is likely to govern its dynamics at longer timescales, especially at the point where the actuation ceases to be efficient. For this reason, the nonlinear effects are essential to determine how effective the actuation may be at suppressing instabilities, and how long it may remain so.

Including the nonlinear dynamics, however, raises several technical difficulties. First, the optimal amplitude and phase were determined without any consideration of nonlinearity, so we may question whether these remain optimal for the nonlinear evolution. The answer lies partly in how the suppression is assessed. Following the approach we adopted in the linear study, we sought the point of lowest amplitude for the leading eigenmode and the time of this minimum. Since the evolution from an infinitesimal perturbation to that point involves only very small amplitudes, it is legitimate to assume that nonlinearities would play little role there and that, consequently, the linear and the nonlinear evolutions would remain very similar during this phase~\citep{drazin_reid_2004}. One should, however, keep in mind that if phase and amplitude were sought so as to optimise the saturated state of the evolution, the full nonlinear equations would need to be included in the optimisation process \citep{Pringle:JFM2012}. However, our specific purpose of understanding the influence of the nonlinear effects is better served by keeping a similar approach to the linear study. Additionally, since the linear study showed that the growth of the leading eigenmode was only effectively suppressed until $t=t_f$, we shall only apply the actuation up to that time, and let the flow evolve freely after that time.

Second, the definition of the initial conditions for the nonlinear problem differs from the linear one: the linear equations indeed return the same evolution regardless of the amplitude of the initial condition (up to a multiplicative factor), and only the relative amplitude and phase of the actuation mattered. To transpose our approach into the nonlinear framework, we shall also specify the amplitude and phase of the actuation relative to the perturbation. In the nonlinear equations, however, if the unstable mode is left to grow ``naturally'' from noise, its amplitude and phase are not specified in the initial condition, so actuation cannot be fixed \emph{a priori}. A workaround would be to let the unstable mode develop up to a trigger-amplitude, where both amplitude and phase can be measured, and then apply the optimal actuation based on these. Indeed, such an approach would be required in a real process where the onset of instability would need to be detected for the actuation to be activated.  For instance, electromagnetic sensors~\citep{thomas2001:ISIJ,Cho2019:Metal} can detect the onset of such instabilities. Since our previous study of the free nonlinear evolution of the leading eigenmode confirmed that it emerged naturally from noise \citep{Kumar:JFM2020}, we shall directly initiate the nonlinear simulation with the leading eigenmode set to an amplitude such its ratio to the base flow
\begin{equation}
\mathcal{R}_A = \frac{({\bf q}^\prime, {\bf q}^\prime)}{({\bf Q},{\bf Q})}
\end{equation}
remains much smaller than unity (see table~\ref{tab:Simulation_details}). Here, ${\bf Q}=({\bf U}, \bar{T})^\top$ represents the steady base flow. Note that for the nonlinear simulations, ${\bf q}^\prime$  is obtained by subtracting the base flow from the entire flow field.

\begin{figure}
\begin{center}
\begin{tikzpicture}[scale = 0.8, every node/.style={transform shape}]
\node[anchor=north, inner sep=0] (image) at (0,0) {\includegraphics[width=\textwidth]{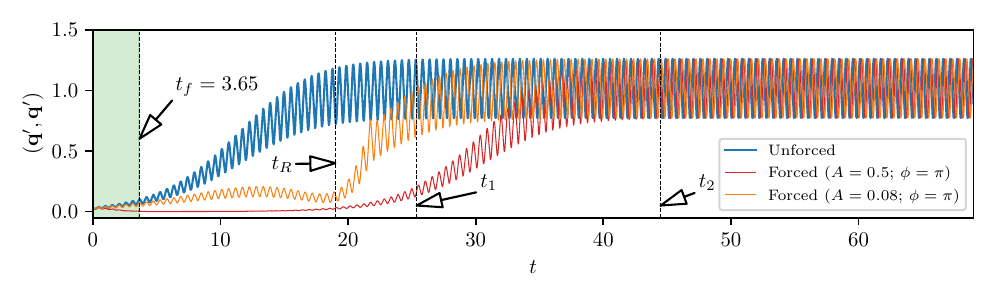}};
\node [below] at (-8.4,0.3) {(a)};
\node[anchor=north, inner sep=0] (image) at (0,-6) {\includegraphics[width=\textwidth]{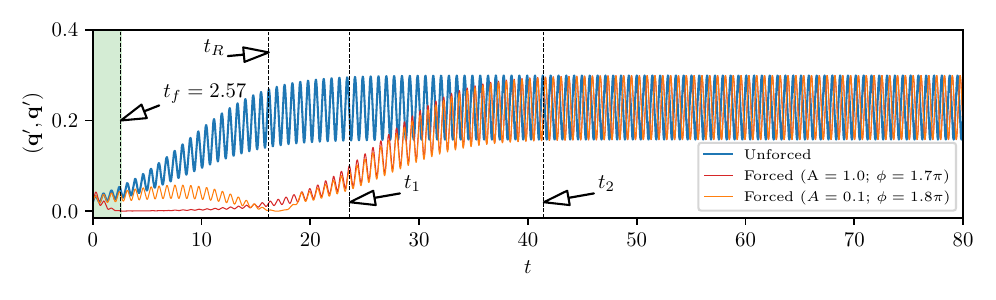}};
\node [below] at (-8.4,-5.6) {(b)};
\node[anchor=north, inner sep=0] (image) at (0,-12) {\includegraphics[width=\textwidth]{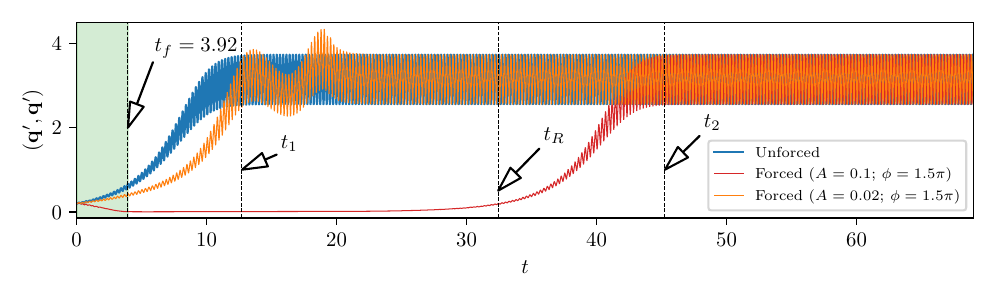}};
\node [below] at (-8.4,-11.6) {(c)};
\node[anchor=north, inner sep=0] (image) at (0,-18) {\includegraphics[width=\textwidth]{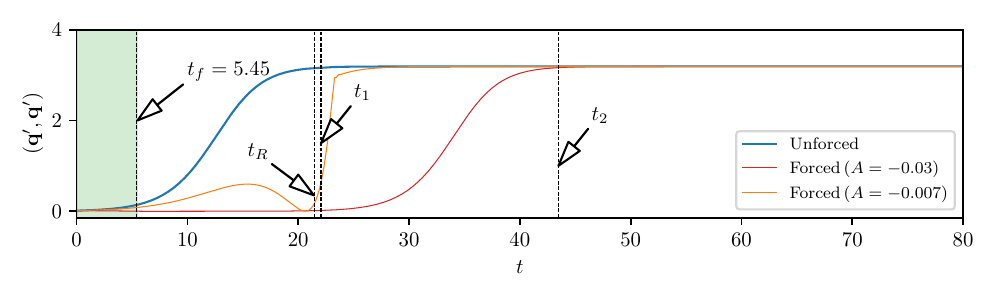}};
\node [below] at (-8.4,-17.6) {(d)};
\end{tikzpicture}
\end{center}
\caption{As a function of time $t$, the nonlinear evolution of the perturbation energy $({\bf q}^\prime, {\bf q}^\prime)$,  in the simulation cases (a) C1, (b) C2, (c) C3, and (d) C4,  without actuation (blue), with actuation maximising $t_f$ (orange) and with actuation achieving maximum reduction of energy (red).} 
\label{fig:Nonlinear_Receptivity_Forcing}
\end{figure}

\subsection{Analysis of cases C1-C4}
The evolution of the perturbation with and without actuation for case C1 is shown in figure \ref{fig:Nonlinear_Receptivity_Forcing}(a). The blue curve illustrates the free nonlinear evolution with the flow initialised with the leading eigenmode for case C1 and no actuation applied. Initially, the energy of the perturbation $({\bf q}^\prime, {\bf q}^\prime)$ grows exponentially, as predicted by the linear theory, and this confirms that nonlinear effects do not affect the early stages. These indeed induce a saturation at $t=t_1=25.4$. The envelope of the curve provides an estimate of the saturation time, denoted as $t_1$. We consider the saturation time to be reached when the derivative of the envelope with respect to time falls below $10^{-3}$. The red curve represents the evolution of the perturbation for the same initial conditions, but this time, with the suppression-optimising actuation applied as we just described. Until $t \approx 10$, the actuation results in a decrease of the perturbation's amplitude, after which the perturbation begins to grow slowly.  Interestingly, the decay of the perturbation continues well beyond the time of minimum amplitude found in the linear simulation. Since, however, nonlinear effects are not expected to play a role for such low amplitudes, as suggested by the exponential form of the actuation-free evolution, the difference more likely results from switching off the actuation at $t_f$ in the nonlinear simulations. Given the very low amplitude of the perturbation around $t=t_f$, the system remains governed by linear dynamics so the reason for the extended period of suppression is that exponential growth from such a low level of perturbation simply extends over a longer period of time. This suggests that applying the actuation beyond the time of maximum growth suppression in fact promotes the growth of the eigenmode. As such, switching off the actuation at $t_f$ is optimal. The underlying suppression strategy, is then to suppress the leading eigenmode down to the lowest possible amplitude so that the linear mechanism takes as long as possible to act. The nonlinear dynamics only provide a measure of when this strategy ceases to be effective.

Indeed, after $t \approx 20$, the exponential growth becomes sufficient to activate nonlinearities and the perturbation amplitude reaches nonlinear saturation at $t=t_2$ ($t_2$ is calculated using the same method as $t_1$ by analysing the envelope of the curve). In order to estimate the effectiveness of the actuation at suppressing the instability, we define the nonlinear efficiency time, $t_{\rm eff}=t_2 - t_1$, which for C1 is $19.1 \simeq 5.23t_f$. In other words, the actuation prevents nonlinear saturation during $5.23$ times the duration of its application, \emph{i.e.} it approximately doubles the time to saturation. Even more interestingly, the unstable mode returns to its initial amplitude at a time $t_R$, $\eta_a=5.21$ times longer than $t_f$, so that it could potentially be indefinitely maintained at that level by applying the actuation during $t_f$ every $\eta_at_f$, which costs $\eta_a$ less energy than applying the actuation continuously.  These results are very encouraging since they validate the linear approach, and show that the receptivity forcing obtained from the linear analysis is indeed capable of suppressing the growth of the instability in the full nonlinear system for a long time, not only compared to the timescale of growth of the instability and to the duration of the actuation $t_f$.

In addition to applying actuation based on the minimum value of $({\bf q}^\prime, {\bf q}^\prime)$, we also tested the actuation strategy based on the forcing maximising $t_f$. For Case C1, the highest $t_f$ value of 19.1 is achieved for $A=0.08$ and $\phi=\pi$. The simulation corresponding to these parameters is depicted by the orange curve in figure~\ref{fig:Nonlinear_Receptivity_Forcing}(a). Interestingly, the perturbation energy remains below that of the unforced case until around $t \approx 20$, yet it consistently stays higher than that of the mode with suppression-optimising actuation ($A=0.5$), before reaching nonlinear saturation. Since the perturbation energy for $A=0.08,\phi=\pi$ exceeds that of $A=0.5, \phi=\pi$, its nonlinear saturation occurs prior to the case of $A=0.5$. This means that suppression-optimising actuation outperforms time-optimising actuation.

Case C2 returned exactly the same phenomenology for the suppression-optimising actuation, despite the difference in the nature of the leading eigenmodes (see figure~\ref{fig:Nonlinear_Receptivity_Forcing}(b)). The values of $t_{\rm eff}/t_f$ and $\eta_a$ are not significantly different than for C1, but the time-optimising actuation incurs a different behaviour: although the unstable mode does not initially decay, it decays briefly at a much later time than for the actuation optimising suppression, down to a similar level. Shortly after, however, both curves practically coincide and simultaneously evolve to saturation. In this sense both actuations perform equally well: This is the only case where the suppression-optimising actuation does not significantly outperform the time-optimising actuation. Nevertheless, since the perturbation remains at a much lower level in the initial stages of evolution for the suppression-optimising actuation, it is still overall better at suppressing the unstable mode.

Case C3 also produced the same phenomenology shown on figure~\ref{fig:Nonlinear_Receptivity_Forcing}(c), but with increased values of $t_{\rm eff}/t_f=8.29$ and $\eta_a=8.27$, \emph{i.e.} with a much higher efficiency of the suppression-optimising actuation. By contrast with C2, however, the time-optimising actuation is practically ineffective at suppressing the unstable mode as its amplitude grows to saturation barely later than without any actuation.

Lastly, we applied constant actuations in the nonlinear simulation of case C4, since the unstable mode is non-oscillatory. Once again, we used the unstable mode as the initial condition.  The non-oscillatory nature of the mode does not seem to play a role as the evolution of the mode's energy in all three simulations, without actuation, with suppression-optimised actuation and with time-optimising actuation show very similar features to C1 and C3, with $t_{\rm eff}/t_f \approx \eta_a=3.93$.

\subsection{The salient features of the nonlinear evolution}
The overall outcome of the nonlinear simulations is that the phenomenology seen on in the  linear simulation remains valid until the nonlinear effects become important and crucially, this happens when the saturation starts. If an actuation based on the adjoint mode, applied up to the time of maximum suppression $t_f$ the unstable mode does not regain its initial amplitude until a time $\eta_a t_f$, roughly an order of magnitude longer than $t_f$ ($\eta_a=8.27$ in the most favourable case C3), so as long as the amplitude and phase of the actuation are optimised for suppression (and not for $t_f$). Until that point, the evolution follows mostly the linear dynamics, as the energy of the unstable mode remains small compared to that of the base flow. The nonlinear effects act shortly after this time, and when they do so, they incur growth up to the same point of saturation as when no actuation is applied. 

In light of this behaviour, an {\em on-off control strategy} could offer a viable control strategy. Repeating a sequence where the suppression-optimising actuation is applied until $t_f$, then switched off to let the flow evolve until $\eta_a t_f$, may indefinitely keep the flow evolution in the linear regime, where the actuation remains effective. Thus, this strategy may confine oscillations to very small amplitudes for an arbitrary length of time. Though, this strategy needs to be further explored to assess how feasible this approach is in practical applications.

\section{Conclusions}
\label{sec:conclusions}
Inspired by the process of continuous casting of liquid metal alloy, we sought to model the suppression of oscillations in mixed baroclinic convection in a nearly hemispherical cavity. This problem offered us an opportunity to assess the feasibility of suppressing oscillatory instabilities by means of an actuation modelled on the receptivity map of unstable modes. 
Doing so led us to consider four canonical cases spanning the three branches of instability for this problem~\citep{Kumar:JFM2020}: a purely convective flow subject to a supercritical oscillatory instability (C1), a mixed convective flow subject to a supercritical oscillatory instability (C2), a mixed convective flow subject to a subcritical oscillatory instability (C3), and a mixed convective flow subject to a supercritical non-oscillatory instability (C4).

For this, we first identified the receptivity map for each of these cases. We found that as the intensity of the inflow increases, the region of receptivity becomes increasingly concentrated near the inlet surface, \emph{i.e.} increasingly favourable in the industrial context where immersing an actuator in the bulk of the melt over extended periods of time is not a feasible option. To gain insight into the instability mechanism, we analysed the sensitivity tensor. We found that for the low Prandtl number of liquid metals, the temperature does not respond to any actuation feedback, whether thermal or mechanical, whereas the velocity field is responsive to both thermal and mechanical actuation. This is understandable since thermal diffusion acts nearly two orders of magnitudes faster than viscous diffusion. As such, acting on the temperature field would require a mechanism $\mathcal{O}(Pr)$ faster than one acting on the velocity field. For C2-C4, the sensitivity map was located in the lower half of the domain, hence physically disconnected from the receptivity map. The disconnect was less obvious in case C1, but still present, despite a slight overlap between the two maps. This indicates that an actuation in the receptivity region would act on the amplitude of the unstable mode but not affect its eigenvalue, which would react to a forcing aligned with the sensitivity map. With these results in hand, we performed linear and nonlinear simulations of the evolution of the unstable mode with and without actuation to answer the four questions set out in the introduction:
\begin{itemize}
\item[(i)]
Since, for all cases (except C1), the area of receptivity is located near the inflow surface, the most efficient way to act on the instability in practice is to modify the inflow. This area is one of the most accessible in casting devices, so this result is favourable to the application. It must, nevertheless, be understood in the context of the simplified model we are considering here, and the question remains open regarding how the receptivity area would change in a more realistic geometry. 
\item[(ii)]
Implementing a thermo-mechanical actuation modelled on the topology of the adjoint to the unstable mode to be suppressed (\emph{i.e.}, its receptivity map) led to a suppression of the unstable eigenmode. To find this result, we scanned possible values of the amplitude and phase of the actuation with respect to the unstable mode.  In doing so, we found that it was always possible to find a combination of amplitude and phase of the actuation leading to a significant suppression of the unstable mode. Typically, its energy could be reduced by two orders of magnitude in all four cases.
\item[(iii)]
If kept at a constant amplitude, the actuation only had a stabilising effect during a finite time $t_f$ of the order of the inverse growth rate of the unstable mode $\sigma^{-1}$, after which linear simulations proved it to be \emph{destabilising}, in the sense that it enhances the growth of the unstable mode, compared to the simulations without actuation. The reason for this rebound is that, once the instability is suppressed, the actuation acts as an extra force driving the system away from the equilibrium point corresponding to the base flow. This led us to consider two strategies for the design of the actuation: one where amplitude and phase of the actuation are chosen to optimise the suppression of the mode's energy (suppression-optimising actuation), and one optimising $t_f$, \emph{i.e.} for which the decay of the unstable mode is the longest (time-optimising actuation). On the other hand, if the amplitude of the actuation was kept proportional to the time-dependent amplitude of the unstable mode all along its evolution, the actuation based on receptivity asymptotically led to the full suppression of the unstable mode. Although this method is more difficult to implement in practice, it demonstrates that an actuation based on the receptivity map effectively suppresses the unstable mode.
\item[(iv)]
To evaluate the influence of nonlinearities and verify if the actuation led to a modification of the base flow and its corresponding equilibrium point, we calculated the evolution of the unstable mode through the full nonlinear governing equations, with either type of actuation applied during $t_f$, after which the flow was left to evolve freely. With suppression-optimising actuation, the energy of the unstable mode always decayed well beyond $t_f$ before it slowly re-grew, to recover its initial amplitude at a time $\eta_a t_f$ typically an order of magnitude greater than $t_f$. In that sense $\eta_a$ measures the efficiency of the actuation.  Time-optimised actuation always led to significantly lower values of $\eta_a$, and even to no suppression at all in case C3. Shortly after $\eta_a t_f$, nonlinearities incurred a rapid growth leading to a saturation at the same level as nonlinear simulations without actuation. In all cases, however, the suppression phase followed the prediction of the linear simulation where the base flow was fixed. This confirmed that the base flow is not measurably affected by the actuation and confirms that actuations based on the receptivity map acts on the unstable mode only.
\end{itemize}

Crucially, these results applied to all four cases, regardless of whether the unstable mode is oscillatory or not, and whether it arises through a supercritical or a subcritical bifurcation. These results open interesting perspectives in view of the suppression of the instability. A possible strategy would consist in applying a mechanical actuation in the inflow region based on the receptivity map of the unstable mode, with amplitude and phase chosen to optimise the suppression of the optimal mode, through the \emph{linear} evolution equations. Actuation should then be applied from the point where the unstable mode reaches a small, arbitrary threshold amplitude, until time $t_f$. For $t>t_f$, the flow should be left to evolve until the unstable mode regains its initial threshold amplitude, at $t=\eta_a t_f$, at which point the actuation should be applied again, iterating the procedure for as long as the flow needs to be stabilised. The success of this strategy relies on the assumption that the threshold amplitude is sufficiently small for the unstable mode to evolve according to the linear equations. In other words, the strategy consists of preventing it from entering a nonlinear regime where the actuation would become ineffective.

The application of this idea raises a number of questions for further studies. First, we considered only a single unstable mode (or at most two). While similar restrictions apply to stabilisation techniques based on the properties of the adjoint equations, it is unclear how far beyond criticality this method would remain effective. Certainly, the cases studied in this paper showed that it remained effective even when more than one mode from the same branch was unstable (as in case C2).  Another novel aspect of the mixed-convective problem we considered here, compared to the classical cylinder wake problem, is that in more supercritical flows, several branches of instabilities with different underlying mechanisms may become unstable. Since, however, the approach is linear, a linear combination of optimal actuations for each of the unstable modes may succeed in stabilising all of them, provided transient growth due to their possible non-orthogonality does not lead to perturbation amplitudes capable of igniting nonlinearities~\citep{schmid2001}. Such an approach may involve the stabilisation of modes from different branches if the flow becomes multi-modal, as for example magnetoconvective flows do~\citep{xu2023_prf}. In case one of these modes is subcritical (as for case C3), there is an additional risk that an altogether different branch of instability may drive the transition, even below the critical Rayleigh number of the linear stability of the leading eigenmode. Nonlinear simulations in the subcritical regime have shown subcritical convection was not ignited by the simple addition of random noise \citep{Kumar:JFM2020}. A similar phenomenon occurs in quasi-two dimensional shear flows \citep{camobreco2021_prf,camobreco2022_jfm}, where it turns out that the subcritical transition is still controlled by the linearly unstable mode. If this is still the case, then a receptivity-based actuation would still be expected to be able to suppress instability in the subcritical regime.

\section*{Acknowledgements}
Part of this research was funded by Constellium Technology Center (C-TEC).} Our numerical simulations were performed on {\em Zeus} of Coventry University and the {\em Sulis} Tier 2 HPC platform hosted by the Scientific Computing Research Technology Platform at the University of Warwick. A. Poth{\'e}rat acknowledges support from EPSRC, under grant No. EP/X010937/1.
%
%
\appendix
\section{Sensitivity tensor \label{sec:sensitivity_tensor}}
\begin{figure}
\begin{center}
\begin{tikzpicture}[scale = 0.8, every node/.style={transform shape}]
\node[anchor=north, inner sep=0] (image) at (0,0) {\includegraphics[width=\textwidth]{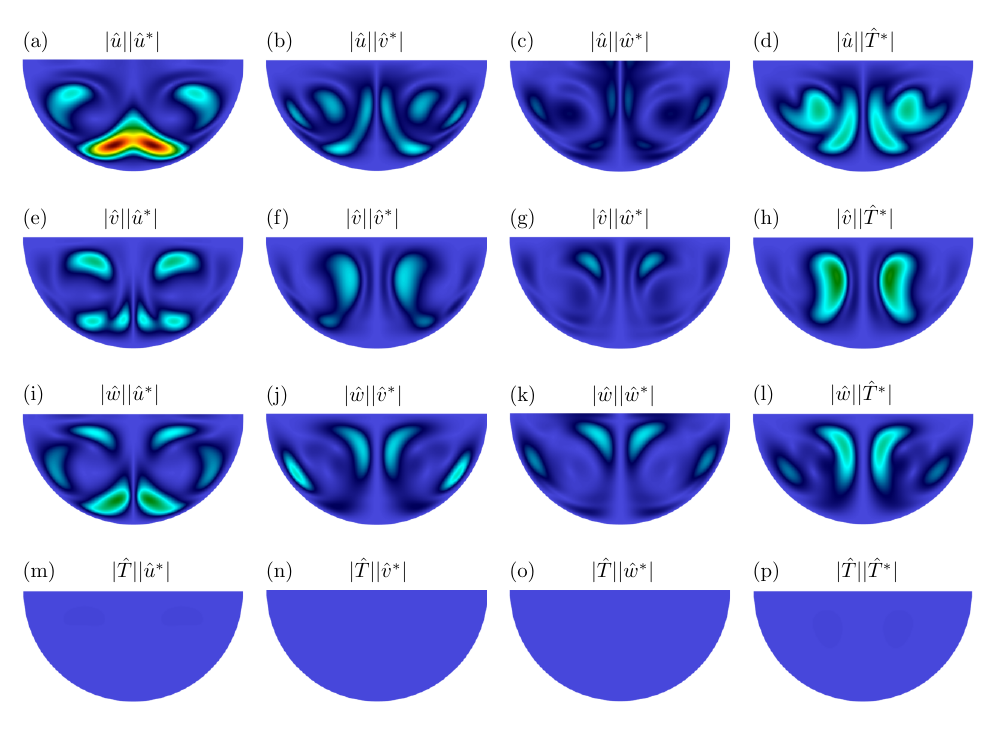}};
\node[anchor=south west,inner sep=0] (image) at (-1,-14) {\includegraphics[width=0.15\textwidth]{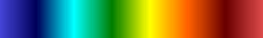}};
\node [below] at (-1,-14) {\scriptsize $0.0$};
\node [below] at (2,-14) {\scriptsize $3.7$};
\end{tikzpicture}
\end{center}
\vspace{-0.5\baselineskip}
\caption{The absolute value of the components of the sensitivity tensor $S_{ij} = \hat{\bf q}_i \hat{\bf q}_j^*$ for the simulation case C1 ($Re=0$,  $Ra=7 \times 10^3$, and $k=6$) in the $x-y$ plane.} 
\label{fig:sensitivity_tensor_Re_0}
\end{figure}
We examine the sensitivity tensor $S_{ij}$ to determine how individual components of the temperature and velocity fields feedback on each other to drive the instability associated to the leading eigenmode.

\begin{figure}
\begin{center}
\begin{tikzpicture}[scale = 0.8, every node/.style={transform shape}]
\node[anchor=north,inner sep=0] (image) at (0,0) {\includegraphics[width=1\textwidth]{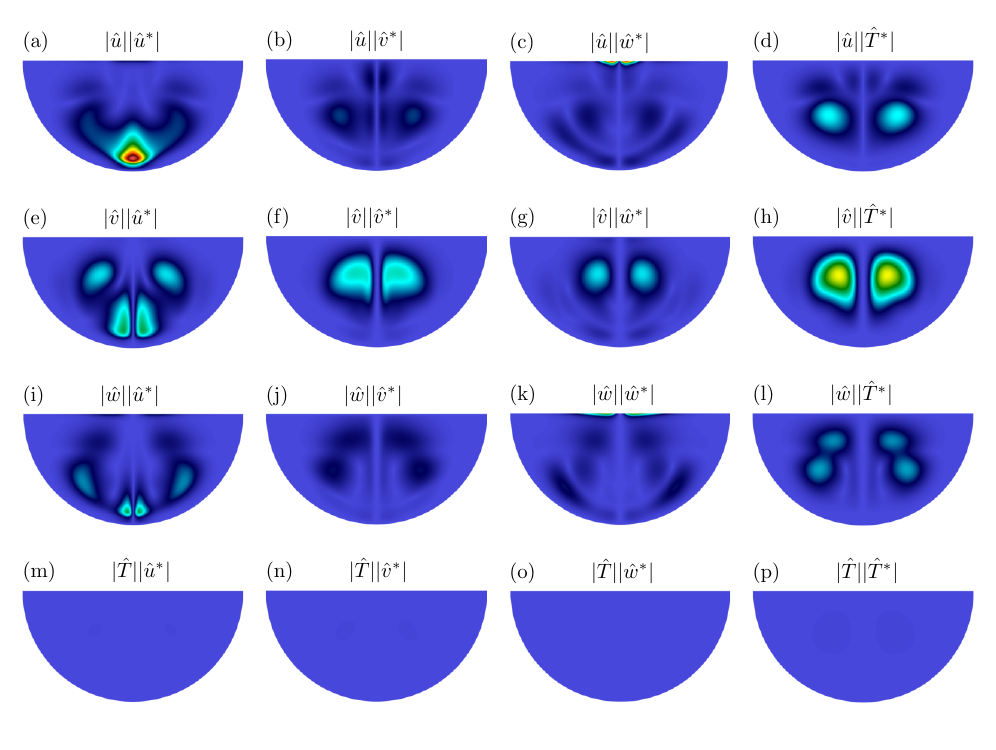}};
\node[anchor=south west,inner sep=0] (image) at (-1,-14) {\includegraphics[width=0.15\textwidth]{fig9_12_colorbar.png}};
\node [below] at (-1,-14) {\scriptsize $0.0$};
\node [below] at (2,-14) {\scriptsize $5.6$};
\end{tikzpicture}
\end{center}
\vspace{-0.5\baselineskip}
\caption{The absolute value of the components of the sensitivity tensor $S_{ij} = \hat{\bf q}_i \hat{\bf q}_j^*$ for the simulation case C2 ($Re=50$,  $Ra=7 \times 10^3$, and $k=6$) in the $x-y$ plane.} 
\label{fig:sensitivity_tensor_Re_50}
\end{figure}
\begin{figure}
\begin{center}
\begin{tikzpicture}[scale = 0.8, every node/.style={transform shape}]
\node[anchor=north, inner sep=0] (image) at (0,0) {\includegraphics[width=\textwidth]{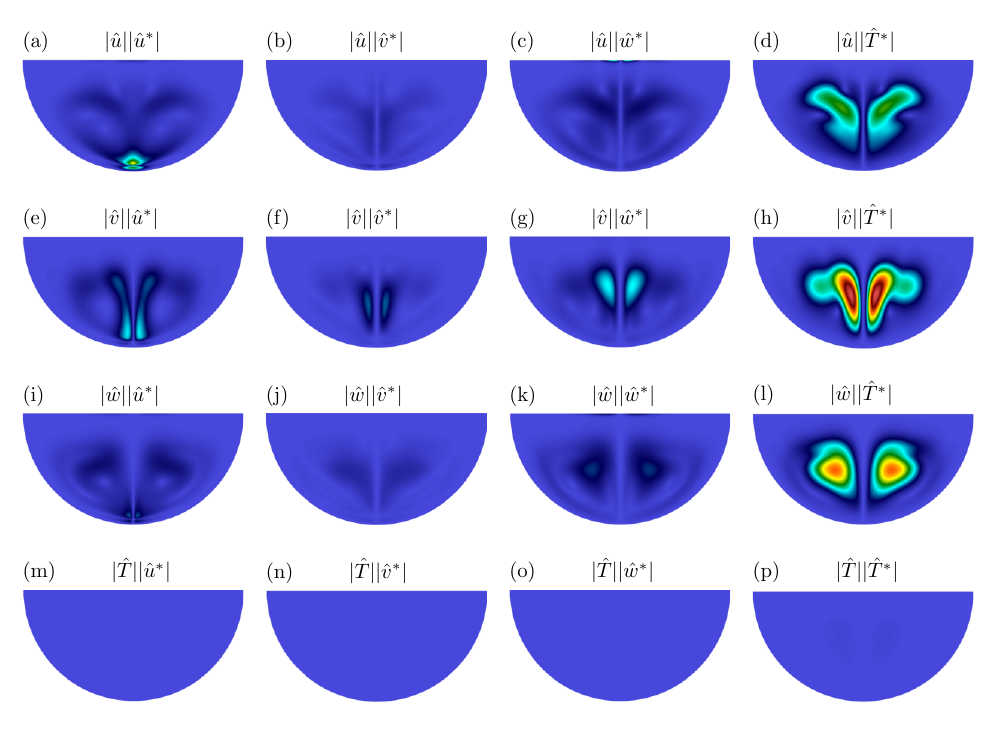}};
\node[anchor=south west,inner sep=0] (image) at (-1,-14) {\includegraphics[width=0.15\textwidth]{fig9_12_colorbar.png}};
\node [below] at (-1,-14) {\scriptsize $0.0$};
\node [below] at (2,-14) {\scriptsize $12.9$};
\end{tikzpicture}
\end{center}
\vspace{-0.5\baselineskip}
\caption{The absolute value of the components of the sensitivity tensor $S_{ij} = \hat{\bf q}_i \hat{\bf q}_j^*$  for the simulation case C3 ($Re=100$, $Ra=4 \times 10^4$, and $k=4$) in the $x-y$ plane.} 
\label{fig:sensitivity_tensor_Re_100}
\end{figure}
\begin{figure}
\begin{center}
\begin{tikzpicture}[scale = 0.8, every node/.style={transform shape}]
\node[anchor=north, inner sep=0] (image) at (0,0) {\includegraphics[width=\textwidth]{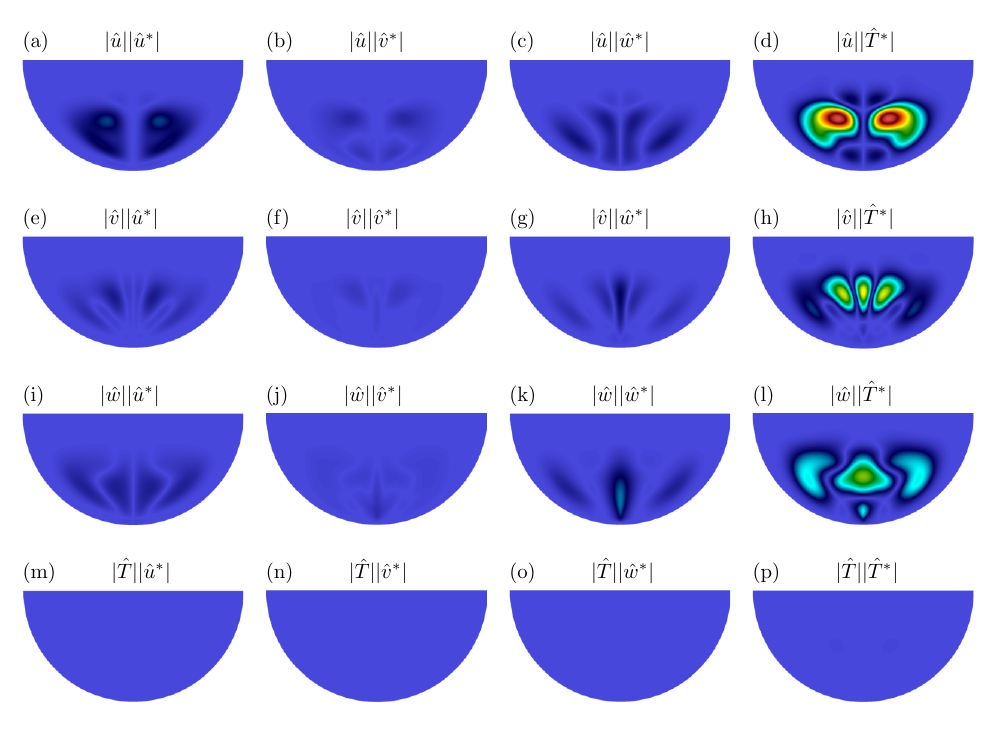}};
\node[anchor=south west,inner sep=0] (image) at (-1,-14) {\includegraphics[width=0.15\textwidth]{fig9_12_colorbar.png}};
\node [below] at (-1,-14) {\scriptsize $0.0$};
\node [below] at (2,-14) {\scriptsize $16.3$};
\end{tikzpicture}
\end{center}
\vspace{-0.5\baselineskip}
\caption{The absolute value of the components of the sensitivity tensor $S_{ij} = \hat{\bf q}_i \hat{\bf q}_j^*$  for the simulation case C4 ($Re=150$, $Ra=8 \times 10^4$, and $k=6$) in the $x-y$ plane.} 
\label{fig:sensitivity_tensor_Re_150}
\end{figure}

The first noticeable feature is that in all four cases C1-C4 represented on figures~\ref{fig:sensitivity_tensor_Re_0} to \ref{fig:sensitivity_tensor_Re_150}, all elements of the tensor associated to $|\hat T|$, are practically identically zero, including $|\hat T||\hat T^*|$. This means that the temperature perturbation responds to feedback from none of the components, not even a thermal one. On the other hand components of the tensor involving a velocity component and  $|\hat T^*|$ all show a strong response, so that the temperature field of the perturbation can likely be affected {by changes in the velocity field}. This can be understood in view of the low value of $Pr$ considered here (and for other liquid metals): even if {mechanical feedback} overcomes viscous forces to act on the velocity field, it would still need to act on a $\mathcal{O}(Pr)$ timescale to overcome thermal diffusion and affect the temperature field. This is because thermal diffusion (low $Pr$) is too fast, preventing thermal feedback from contributing to instability. Consequently, at low $Pr$, the instability relies on feedback through its velocity field, \emph{i.e.}, mechanical.

In case C1, figure~\ref{fig:sensitivity_tensor_Re_0}(a) highlights that the greatest feedback is received by $x$-component of the velocity from the perturbation, from the $x$-velocity component of the actuation. The maximum coincides with the lower part of the flow domain where the baroclinic jets turns up, just  upstream of the point where they meet each other, and where the instability starts.  C2 also features maximum feedback from $\hat u_x^*$ onto $\hat u_x$. This time, however, the sensitivity area is localised right at the bottom of the domain, at the point where the instability is maximum in velocity amplitude. Cases C3 and C4 by contrast, exhibit the greatest response to a temperature feedback. All three components show high response in both cases. In C3, the maximum receptivity is on the $v$ component, but on the $u$ component for C4.


\end{document}